\documentclass[a4paper,11pt]{article}
\pdfoutput=1 

\usepackage{jheppub}                
\usepackage[all]{hypcap}            
\usepackage{slashed}                
\usepackage[usenames,table]{xcolor} 
\usepackage{multirow}               
\usepackage{booktabs}               
\usepackage{empheq}                 
\usepackage[theorems]{tcolorbox}    
\usepackage{feyndiag}               
\usepackage{parskip}                
\usepackage[section]{placeins}      

\graphicspath{ {../plots/} {plots/} }

\newcommand{\tr}{\textrm{tr}}
\renewcommand{\Re}{\operatorname{Re}}

\definecolor{ered}{RGB}{127,0,0}
\definecolor{egreen}{RGB}{0,127,0}
\definecolor{eblue}{RGB}{0,0,127}
\newcommand{\eqred}[1]{{\color{ered}#1}}
\newcommand{\eqgreen}[1]{{\color{egreen}#1}}
\newcommand{\eqblue}[1]{{\color{eblue}#1}}
\definecolor{boxgray}{RGB}{240,240,240}
\newcommand{\eqbox}{\tcboxmath[colback=boxgray,colframe=boxgray,left=1pt,right=1pt,top=2pt,bottom=2pt]}

\title{\textbf Four-Quark Effective Operators at Hadron Colliders}
\author{Maikel de Vries}
\affiliation{DESY Theory Group\\Notkestra{\ss}e 85, 22607 Hamburg, Germany}
\emailAdd{maikel.devries@desy.de}
\preprint{\small DESY 14-158}
\arxivnumber{1409.4657}

\abstract{
The robustness of translating effective operator constraints to BSM theories crucially depends on the mass and coupling of BSM particles. This is especially relevant for hadron colliders where the partonic centre of mass energy is around the typical energy scales of natural BSM theories. The caveats in applying the limits are discussed using $Z'$ and $G'$ models, illustrating the effects for a large class of models. This analysis shows that the applicability of effective operators mainly depends on the ratio of the transfer energy in the events and the mass scale of the full theory. Moreover, based on these results a method is developed to recast existing experimental limits on four-quark effective operators to the full theory parameter space.
}

\begin{document}

\maketitle
\flushbottom

\newpage 

\section{Introduction}
\label{sec:intro}
The first runs of the LHC at $7$ and $8\; \textrm{TeV}$ have not provided us with any signal for new physics beyond the Standard Model. However, these runs have proven to be very effective in excluding regions in parameter space of many BSM theories using direct searches. On the other hand, model independent methods like effective operators have been used to set limits on the same parameter spaces. For example, fermionic contact interactions have been probed in dijet events by ATLAS \cite{Collaboration:2010eza,Aad:2011aj,ATLAS:2012pu} and CMS \cite{Khachatryan:2010te,Khachatryan:2011as,Chatrchyan:2012bf,Chatrchyan:2013muj}. For evaluating the full exclusion potential from LHC, limits from direct searches and effective operator bounds need to be combined for BSM theories. The translation of effective operator limits to parameter regions of BSM theories is the subject of this study.

The experimental analyses consider a set of effective operators and set upper limits on the size of their coefficients, parametrising the deviation from the Standard Model. Typically the coefficient is written as $2 \pi / \Lambda^2$ for dimension six operators and a lower bound on $\Lambda$ is quoted. BSM theories generically have heavy particles that generate effective operators of the types constrained by experiment when integrated out \cite{Buchmuller:1985jz,Grzadkowski:2010es}. To translate the bound on the effective operator to the full theory --- the BSM theory --- two ingredients are necessary: the analytic expression for the effective operator in terms of full theory parameters and the domain of validity for the effective theory. Then exclusion limits for the parameter space of the full theory can be derived from the experimental results.

As a benchmark for this analysis the most explored channel for contact interactions at the LHC is used: the dijet angular analysis constraining four-quark operators of dimension six. Beyond the Standard Model theories that can be constrained by limits on four-quark operators are typically strongly coupled models. These models contain particles similar to the heavy partners of the $Z$ boson or the gluon, known as $Z'$ or $G'$ bosons. A non-exhaustive list contains colour octets from compositeness \cite{Contino:2006nn,Redi:2011zi,Redi:2013eaa}, flavoured $Z'$ models \cite{Buras:2014sba} and explanations for the top forward backward asymmetry \cite{Aguilar-Saavedra:2014kpa} using axigluons \cite{Jung:2014gfa}. In appendix \ref{sec:toymodel} a toy model is constructed based on $Z'$ or $G'$ models. The relevant parameters are the mass of the particle and the coupling strength to quarks. For these toy models the width of the $Z'$ and $G'$ depends solely on the mass and the coupling strength, therefore not introducing any additional parameters. However, for other BSM theories this may be different and the width must be considered independently.

The translation of effective operator bounds to BSM theories is an important method to constrain full theory parameter spaces. In this study the errors made in the aforementioned translation are quantified and are connected with the kinematic parameters of the experiment and the theoretical model. An important quantity is the effective theory expansion parameter which is the ratio of the transfer energy in the events and the mass scale of the full theory. The non-negligible effect of this expansion parameter on the exclusion regions in the full theory parameter space is scrutinised. Conclusively, it is shown that these effects are crucial and should be taken into account. 

\paragraph{Outline} \mbox{} \\
This work is based on a toy model which is described in appendix \ref{sec:toymodel} and the relevant cross sections calculated in appendix \ref{sec:dijetxsec}. These details are not needed for a basic understanding of the work, but are added to ease understanding and usage of the results. First some general aspects of effective operators at hadron colliders are discussed in section \ref{sec:effoperators}. Then in section \ref{sec:analysis} the existing experimental analyses for constraining four-quark effective operators are reviewed and applied to the toy models. For these analyses the exclusion potential is compared between the full and the effective descriptions of the toy models in section \ref{sec:results}. Finally, in section \ref{sec:conclusion} conclusions are drawn and recommendations are made for using effective operators at hadron colliders.

\section{Effective Operators}
\label{sec:effoperators}
In this section effective operators are discussed in general. First hadron colliders are discussed, identifying which kind of effective operators might be constrained. After that beyond the Standard Model physics is connected to these operators, justifying a certain class of toy models. Effective field theories only work at low energies compared to the energy scale of the full theory. The errors introduced in the effective approach are quantified by an expansion in energy scales, which forms the basis of the work. This section is then concluded with a first comparison between the full and effective theory description, when the translation of effective theory limits to the full theory parameter space is discussed.

\subsection{Bounds from Hadron Colliders}
\label{sec:effoperators:hadroncolliders}
For an analysis of constraining effective operators at a hadron collider it is first useful to make the comparison with lepton colliders. Lepton colliders are generally known for their very precise measurements and therefore harsh limits on precision observables and effective operators. Precise measurements and high luminosities lead to strong limits on effective operators compared to the centre of mass energy of the collider. For example, the limit from LEP for the four-fermion operator $eedd$ equals $26 \; \textrm{TeV}$ \cite{Beringer:1900zz}. Hadron colliders are very different, first of all composite particles like protons are being collided and therefore not all and also an unknown amount of the centre of mass energy of the collider is passed to the partons. These partons --- quarks and gluons --- then interact to produce mostly hadronic final states, presenting another source of imprecision. However, what hadron colliders lack in precision they compensate in centre of mass energy. Hence, they possibly provide a source for constraining effective operators to high energy scales, as well. 

The essential difference when looking at effective field theories in both types of colliders is the difference in energy scales between the limits set on the operators and the processes involved at the collider. For a lepton collider nowadays the centre of mass energy is typically around $250 \; \textrm{GeV}$ and the limits reach up to more than $10 \; \textrm{TeV}$. The energy scale of the full theory behind the effective theory must roughly be in the same ballpark as the limits on the effective theory. The reasoning being that full theories operating at lower energy scales would have been excluded by these limits. Therefore, we know that the effective theory provides a good description of the physics at centre of mass energy at a lepton collider. For a hadron collider the typical centre of mass energies are around $10 \; \textrm{TeV}$, resulting in possible partonic centre of mass energies around $2$ to $4 \; \textrm{TeV}$. The typical limits set by the LHC --- the most energetic hadron collider --- are around $10 \; \textrm{TeV}$. We see that the scale separation is much lower\footnote{This issue is even more urgent if we take into account that the typical scales of BSM physics range from $1$ to $5 \; \textrm{TeV}$.} and the validity of the effective description should be subject to investigation.

In hadron colliders usually protons or antiprotons are collided and these collisions produce a range of Standard Model particles. However, the range of particles is severely dominated by QCD production and therefore jet final states, which are hadronised light quarks or gluons. Therefore, if we are looking into what kind of effective operators can be constrained by hadron colliders, the first that come to mind are those involving quarks or gluons. Indeed, from the dimension six operators that parametrise BSM physics \cite{Buchmuller:1985jz,Grzadkowski:2010es}, the most investigated effective operators at the LHC are four-quark operators of the type
\begin{empheq}[box=\eqbox]{equation} \label{eq:effoperators:fourquarkoperators}
	\frac{2 \pi \zeta}{\Lambda^2} \left( \bar{q}_L \gamma^\mu q_L \right) \left( \bar{q}_L \gamma_\mu q_L \right) .
\end{empheq}
In here $\zeta = \pm 1$ accounts for destructive and constructive interference, respectively, and $\Lambda$ is the energy scale of the effective theory. The scope of this article is limited to four-quark operators. These operators form a direct contribution to the dijet cross section $p p \to j j$ at hadron colliders. Then, if one measures distributions of dijet cross sections at hadron colliders, these can be compared with theoretical predictions for the background (QCD) and the signal (effective operators). The comparison, in absence of any deviations from the background, then leads to exclusion limits on coefficients of the effective operators. 

The experimental collaborations ATLAS \cite{Collaboration:2010eza,Aad:2011aj,ATLAS:2012pu} and CMS \cite{Khachatryan:2010te,Khachatryan:2011as,Chatrchyan:2012bf,Chatrchyan:2013muj} have been pursuing this strategy and have set limits on the effective operators like the one in equation \eqref{eq:effoperators:fourquarkoperators}. Currently, the highest limits are set by CMS from analysing the $p_T$ spectrum of the leading jet \cite{Chatrchyan:2013muj}. These limits are 
\begin{equation} \label{eq:effoperators:explimits}
	\Lambda^+ = 9.9 \; \textrm{TeV} \; \textrm{and} \; \Lambda^- = 14.3 \; \textrm{TeV}
\end{equation}
for destructive and constructive interference, respectively. Although not relevant for this work, the experimental collaborations also constrain effective operators using monojet plus missing transverse energy final states \cite{ATLAS:2012zim,CMS:rwa}. These analyses constrain operators consisting of two quarks and two invisible particles, and are relevant for dark matter searches. The validity of the effective description for these experimental results has been discussed in a series of papers \cite{Busoni:2013lha,Busoni:2014sya,Busoni:2014haa,Buchmueller:2013dya} and has been compared to specific models in \cite{An:2013xka,Papucci:2014iwa,Buchmueller:2014yoa}. Moreover, in the Higgs sector similar analyses have been performed in references \cite{Englert:2014cva,Biekoetter:2014jwa}.

\subsection{BSM Physics}
\label{sec:effoperators:bsmphysics}
In general, new physics beyond the Standard Model produces quarks rather than gluons, so in that sense the four-quark operator already matches topologies in BSM physics. Generically, strongly coupled theories are susceptible to effective operator limits, due to their relatively large couplings. High values for the couplings of new resonances to quarks automatically generate large effective operators coefficients. Moreover, the parameter space of these models can not be fully probed by direct resonance searches. A fact caused by the large couplings of these particles, making them very wide and reducing the effectiveness of resonance searches. Therefore, effective operators are a vital method to constrain strongly coupled BSM models. 

For example, in composite Higgs models with partial compositeness, Standard Model quarks are a mixture of elementary and composite quarks. Some flavour implementations allow for large mixing with the composite sector and then the SM quarks have large couplings to a heavy partner of the gluon --- in these models called the colour octet \cite{Contino:2006nn,Redi:2011zi,Redi:2013eaa}. The colour octet --- being sufficiently heavy --- can be integrated out to obtain a four-quark effective operator. Analogously, models explaining the Standard Model flavour using $Z'$ bosons lead to the same four-quark effective operator \cite{Buras:2014sba}. Another example is the introduction of an axigluon to explain the top forward-backward asymmetry \cite{Jung:2014gfa}. This model predicts a new resonance, which when integrated out produces the four-quark operator as well. Finally, these operators can constrain the dark matter to mediator coupling \cite{Dreiner:2013vla}, using the fact the mediator must couple to quarks for significant dark matter production in monojet plus missing transverse energy searches. 

In summary, typically strongly coupled BSM theories predict bosonic resonances with couplings to Standard Model quarks. These resonances are in most cases heavy copies of the electroweak gauge bosons or the gluon. For that purpose two toy models are introduced: a $Z'$ boson which is a heavy partner of the $Z$ boson and a $G'$ boson which is the gluon's partner. Both partners couple universally to the Standard Model quarks governed by a single coupling constant\footnote{Universal couplings to Standard Model quarks is of course not a general feature of BSM physics and depends heavily on the flavour implementation. However, for the purpose of determining the validity of the effective description simplicity prevails over completeness.}. This coupling constant $g$ and the mass $m$ are the fundamental parameters of the model, the details for both toy models can be found in appendix \ref{sec:toymodel}. The coefficients of the effective operators corresponding to the full theory are obtained in section \ref{sec:toymodel:operators} and also depend on $m$ and $g$. Then when translating the experimental limits on effective operators to the full theory it is most conveniently done in the mass versus coupling plane, since this allows for a direct interpretation in many BSM models. Here the focus is on the validity of the EFT description and not in particular on constraining $Z'$ and $G'$ bosons, see references \cite{Dobrescu:2013cmh,LaRochelle:2014tba} for constraints from LHC on these types of models.

\subsection{EFT Expansion}
\label{sec:effoperators:eftexpansion}
An effective field theory is the low-energy description of some full theory with heavy particles. The effective description is in general valid if it describes processes involving energies much smaller than the energy scale of the full theory. This energy scale of the full theory is determined by the masses of the particles in that theory. The higher dimensional operators in the effective theory are obtained if heavy particles in the full theory are integrated out. This can be done through diagrammatic matching and a detailed example is given in appendix \ref{sec:toymodel:operators}. Generically in the full theory the propagators of the massive particles are expanded around zero transfer momenta $q = 0$ to obtain the EFT expansion
\begin{empheq}[box=\eqbox]{equation} \label{eq:effexpansion:expansion}
	\frac{g^2}{q^2 - m^2} = - \frac{g^2}{m^2} \left[ 1 + \frac{q^2}{m^2} + \mathcal{O} \left( \frac{q^4}{m^4} \right) \right] .
\end{empheq}
A coupling $g$ has been introduced and the particle in the full theory has mass $m$. It is shown in appendix \ref{sec:toymodel:operators} and specifically in equation \eqref{eq:toymodel:effcalc:propexp} that the width of the particle does not play a role if the transfer energy $q^2$ goes to zero. The first term in the expansion will be the coefficient in front of a dimension six operator and the other terms in the expansion will be the coefficients for higher dimensional operators involving derivatives.

In the EFT expansion from equation \eqref{eq:effexpansion:expansion} $q^2$ is the energy transferred by the heavy particle in the diagram. For four-quark operators that can be in all channels, so $q^2 = \hat{s}$, $\hat{t}$ or $\hat{u}$. Usually the EFT description is considered valid or applicable if $q^2$ is smaller than $m^2$, since then a converging series is ensured. However, experiments only probe the leading order operator and are neglecting terms of the order of $q^2/m^2$. This introduces large errors when translating back from effective to full theory if the energy at which the experiments operate are close to the mass scale of the full theory.

\begin{figure}[!ht]
	\centering
	\includegraphics[width=0.6\textwidth]{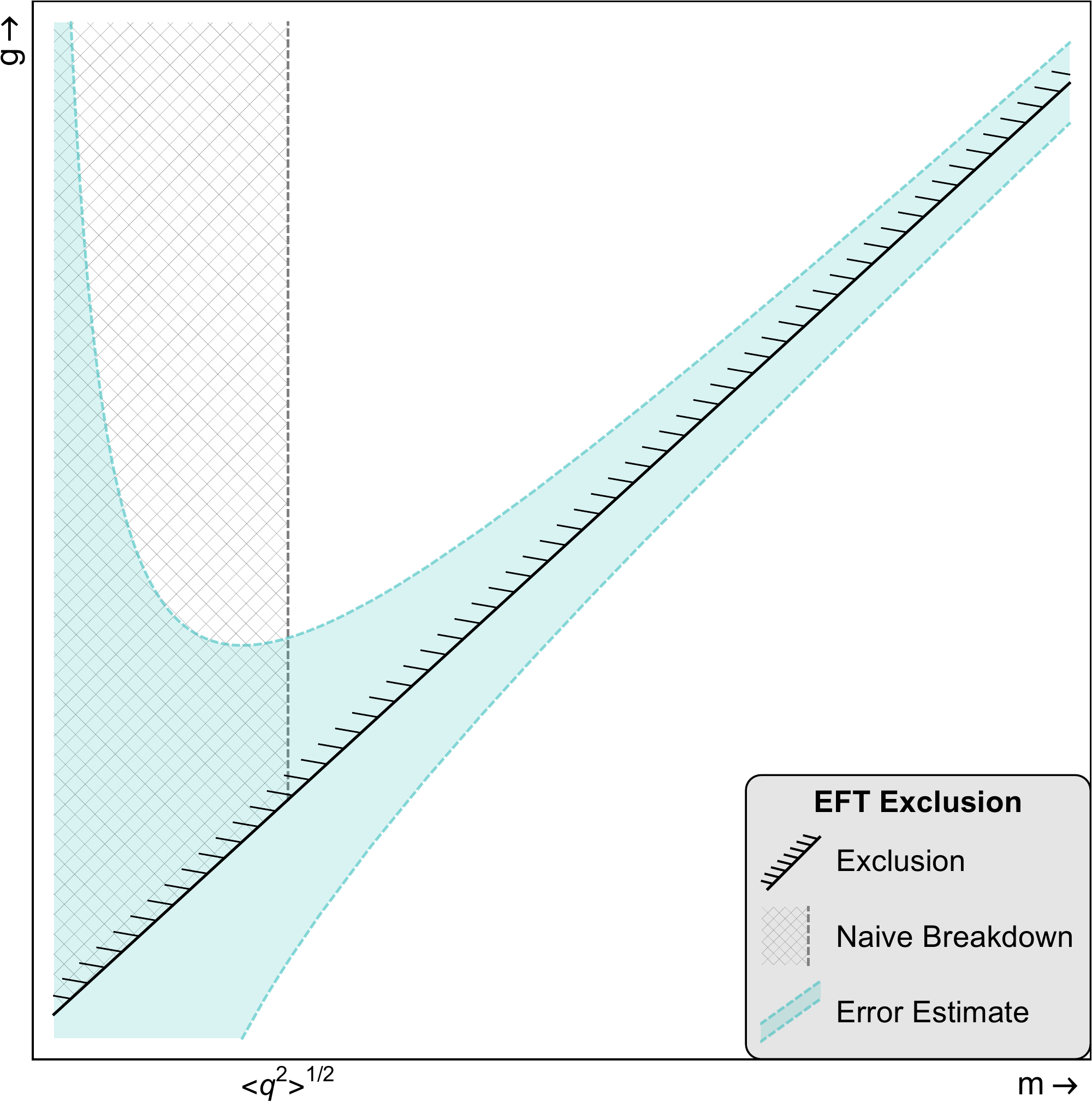}
	\caption{Estimation of the exclusion potential of effective operators in the full theory mass versus coupling parameter space. The effective theory description naively breaks down in the part of parameter space covered by the dashed grey region, where $m^2 < \left\langle q^2 \right \rangle$. In reality the error of the effective description is given by the light blue area which scales as $\left\langle q^2 \right \rangle / m^2$. This figure is just an indication of the effects and actual results are derived in later sections.}
	\label{fig:effoperators:efftofull}
\end{figure}

The experimental results, in the absence of new physics, constrain dimension six operators like in equation \eqref{eq:effoperators:fourquarkoperators}. Comparing these limits to the coefficient in front of the effective operator will constrain the full theory parameter space
\begin{equation}
	\frac{g^2}{m^2} < \frac{2 \pi}{\Lambda^2} .
\end{equation}
A graphical representation of this limit is given in figure \ref{fig:effoperators:efftofull} and the sign indicating interference effects has been absorbed in $\Lambda$. It is to be noted that the naive EFT limit constrains a region above a straight diagonal line in the mass versus coupling plane. Naively the effective description is valid if $m^2 > \left\langle q^2 \right \rangle$, however, more realistically the EFT limit will have an error which scales as $\left\langle q^2 \right \rangle / m^2$. The realistic exclusion can only be obtained by performing the actual analysis and depends on whether the effective theory is over or underestimating the cross section relevant for the analysis compared to the full theory.

For hadron colliders, however, due to the composite nature of the proton the transfer energy is not an exactly known quantity and is not the same for all events used in the experimental analysis. The transfer energy depends on the kinematic requirements of the analysis, see \cite{Busoni:2013lha,Busoni:2014sya,Busoni:2014haa} for a detailed discussion. Therefore, in an experimental context the average transfer energy should be used as a measure for the expansion parameter. This average is an analysis dependent quantity and further discussion is postponed until section \ref{sec:results:eftexpansion}. To allow for a good estimation of the validity of the effective description it is recommendable that experimental analyses quote the average transfer energies in the events used for setting limits on effective operators\footnote{A similar recommendation has been proposed in reference \cite{Englert:2014cva} where running and mixing effects for effective operators have been discussed.}.

Beyond the Standard Model theories may predict $Z'$ or $G'$ like particles which have additional couplings beyond the usual couplings to SM quarks. These additional couplings may be to other heavy particles and increase the width of the $Z'$ or $G'$ like particle. However, these couplings will not affect the production cross section of the dijet final state considered in the experimental analyses. Hence, the effect of these couplings is solely through an additional contribution to the width of the particle, which can be parametrised as $\Gamma_\textrm{BSM}$. Then, if we look at the scaling of $\Gamma_\textrm{BSM}$ with the transferred momentum $q^2$ it is expected to have the same scaling as in equation \eqref{eq:toymodel:dependentwidthnomass} with $m_q$ replaced by the mass of the heavy particle decayed into. From this it immediately follows that also the effect of additional widths can be neglected when looking at the first term of the effective operator expansion, as $q^2$ goes to zero.

\section{Experimental Analyses}
\label{sec:analysis} 
In this section the analyses for obtaining limits on the four-quark effective operators are discussed. Both the full and the corresponding effective theory are analysed according to the ATLAS and CMS prescriptions in order to find out the differences in exclusion potential. Therefore, the experimental analyses are discussed first and then the theoretical application to the limit setting is reviewed in the next section. The understanding of the experimental analyses begins with the calculation of differential dijet cross sections for QCD, the full theory and the effective theory. These cross sections are calculated differentially with respect to $\hat{t}$ in appendix \ref{sec:dijetxsec} and we base this analysis on
\begin{equation}
	\frac{d\sigma}{d\hat{t}} \left( \hat{s}, \hat{t}, \hat{u}, \alpha \right) ,
\end{equation}
where $\alpha$ denotes the collection of the relevant theory parameters for either QCD, the full theory or the effective theory. The results in appendix \ref{sec:dijetxsec} are obtained at leading order in $\alpha_s$, however, next-to-leading order QCD corrections are important as well \cite{Gao:2010bb,Gao:2011ha,Gao:2012qpa}. Unfortunately, inclusion of these effects is beyond the scope of this work, since the focus is on the validity of the effective field theory expansion. In the experimental setting, the partonic cross sections need to be transformed to realistic cross sections using parton density functions. Moreover, to apply kinematic cuts, the cross sections should be differential in certain kinematic variables. These steps are discussed in the rest of this section for the different experimental analyses.

\subsection{Differential Cross Sections}
\label{sec:analysis:differentialxsec}
For four-quark effective operators there have been two types of analyses to date at the LHC: dijet angular distributions \cite{Collaboration:2010eza,Aad:2011aj,ATLAS:2012pu,Khachatryan:2010te,Khachatryan:2011as,Chatrchyan:2012bf} and leading jet $p_T$ spectrum \cite{Chatrchyan:2013muj}. The first type and the necessary kinematics are discussed in this section. However, the first step from partonic cross sections to an actual analysis in a hadron collider is folding with parton density functions. For the partonic cross sections differential in $\hat{t}$ the identification
\begin{equation} \label{eq:analysis:partonicxsec}
	\frac{d^3 \sigma}{d x_1 d x_2 d \hat{t}} (pp \to 34) = f_1(x_1) \, f_2(x_2) \frac{d \sigma}{d \hat{t}} \left( 12 \to 34 \right)
\end{equation}
gives the full cross section. In this formula $12 \to 34$ denotes the partonic process and $x_1$, $x_2$ are the momentum fractions for partons $1$ and $2$. However, this is still differential in $\hat{t}$ and not in the variables used in experiments like the rapidity of the dijet system $Y = \tfrac{1}{2} (y_3 + y_4)$ and the invariant mass of the dijet system $m_{jj}^2 = \hat{s}$. The momentum fractions in terms of these variables are
\begin{equation} \label{eq:analysis:momentumfrac}
	x_1 = \sqrt{\frac{\hat{s}}{s}} \, e^Y \quad x_2 = \sqrt{\frac{\hat{s}}{s}} \, e^{-Y} ,
\end{equation}
where $s$ is the centre of mass energy of the $pp$ collider. From this the differential cross section in terms of the variables defined previously is derived to be
\begin{equation} \label{eq:analysis:differentialxsec}
	\frac{d^3 \sigma}{d Y d \hat{s} d \hat{t}} = x_1 f_1(x_1) \, x_2 f_2(x_2) \frac{d \sigma}{d \hat{t}} \frac{1}{\hat{s}} .
\end{equation}
The integration limits on $Y$ and $\hat{s}$ are determined by the individual experimental analysis and the variable $\hat{t}$ might still be converted to an experimental observable. Note that the limits on $Y$ are also influenced by the limits on the momenta fraction $0 < x < 1$, which give
\begin{equation} \label{eq:analysis:ylimits}
	\left| Y \right| < \frac{1}{2} \log \frac{s}{\hat{s}} .
\end{equation}
By construction the partonic centre of mass energy is limited by the collider's centre of mass energy $\hat{s} < s$, providing an upper limit for the $\hat{s}$ integration.

\subsection{Angular Distribution}
\label{sec:analysis:angulardist}
\begin{figure}[!ht]
	\centering
	\includegraphics[width=0.46\textwidth]{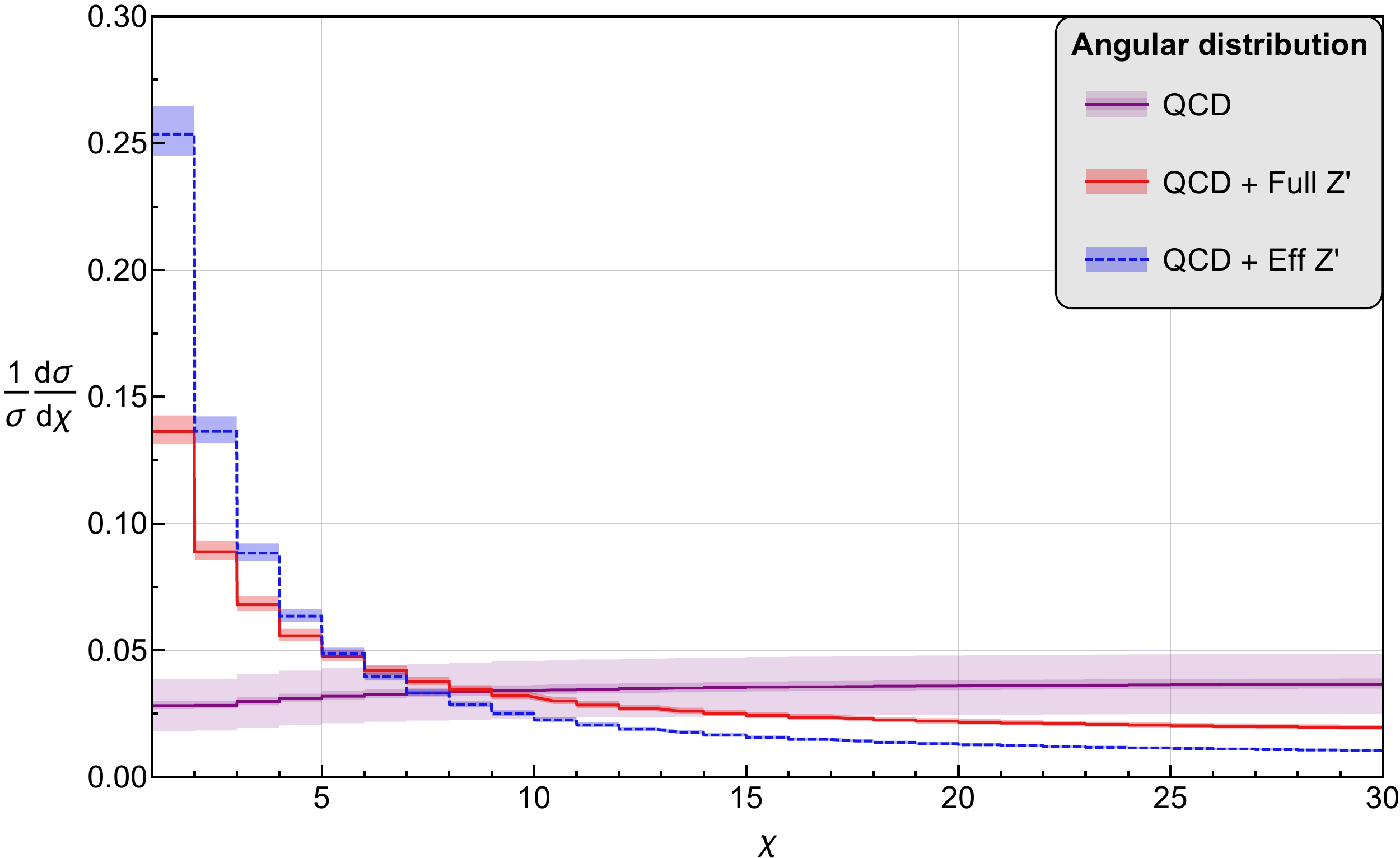} \hspace{2mm}  
	\includegraphics[width=0.46\textwidth]{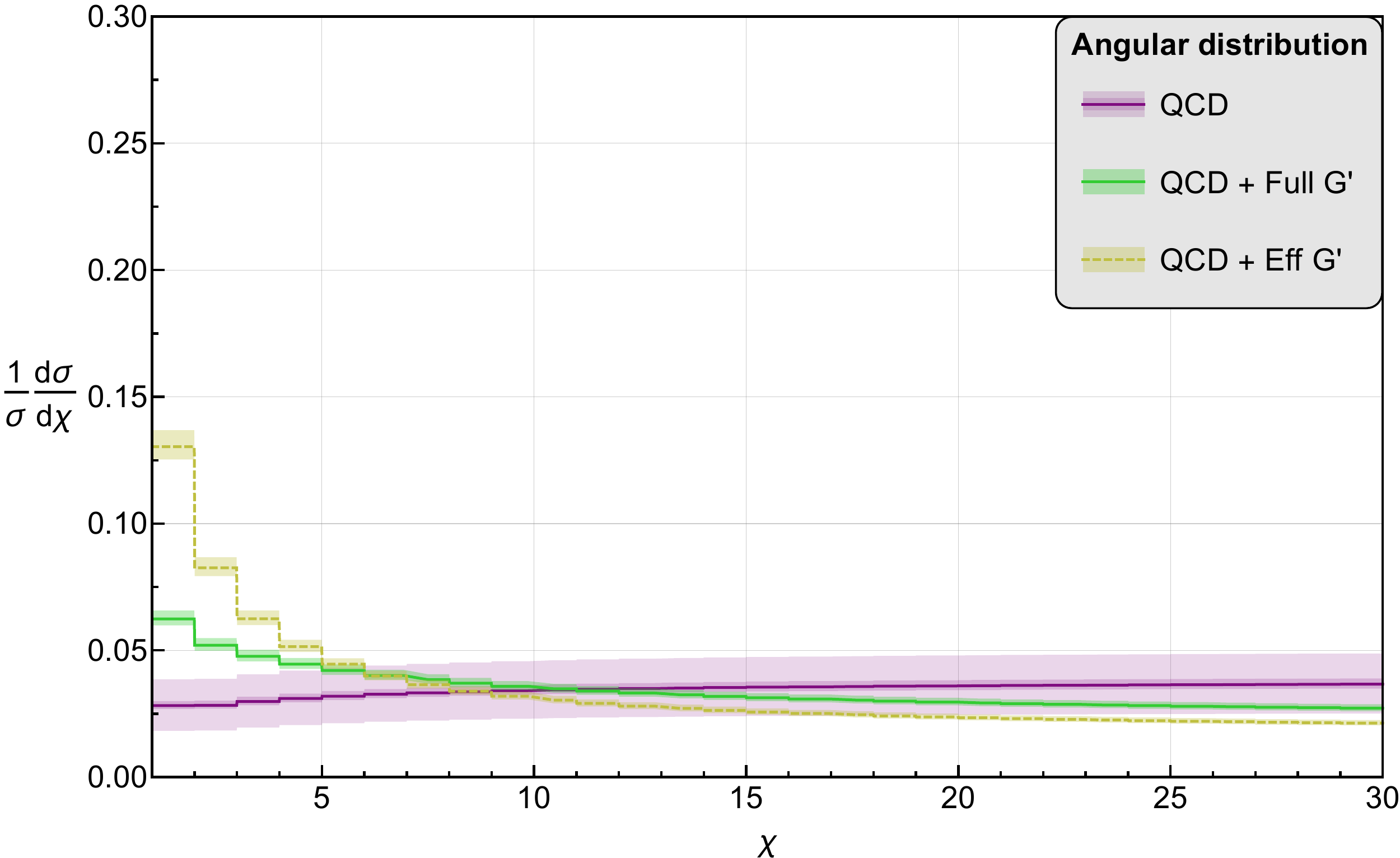} 
	\caption{Reconstruction of the experimental angular distributions for QCD, for the $Z'$ model (left) and the $G'$ model (right) with parameters $m_{Z'} = m_{G'} = 2 \; \textrm{TeV}$ and $g_{Z'} = g_{G'} = \tfrac{\pi}{2}$. This distribution has been obtained for the centre of mass energy integration from $\sqrt{\hat{s}_\textrm{min}} = 3 \; \textrm{TeV}$ to $\sqrt{\hat{s}_\textrm{max}} = 5 \; \textrm{TeV}$. The bands around the different distributions represent for QCD the theory error (inner band) and statistical error (outer band). For the full and effective theory the bands represent the theory error, for which more details are given in section \ref{sec:analysis:erroranalysis}.}
	\label{fig:analysis:dsigmdchi}
\end{figure}

In the CMS analyses \cite{Khachatryan:2010te,Khachatryan:2011as,Chatrchyan:2012bf} based on the angular distribution, events are selected by cuts on the total rapidity of the system $Y$ and are grouped in bins of invariant mass $\hat{s}$. This can be reconstructed by integrating equation \eqref{eq:analysis:differentialxsec} over these kinematic variables. The remaining data is then binned in the variable
\begin{equation} \label{eq:analysis:chi}
	\chi \equiv e^{\left| y_3 - y_4 \right|} = - \left( 1 + \frac{\hat{s}}{\hat{t}} \right) ,
\end{equation}
which represents the angular distribution of the dijet system. It is therefore necessary to obtain the cross section differential in $\chi$ rather than $\hat{t}$. Calculating the Jacobian from equation \eqref{eq:analysis:chi} --- finding $\tfrac{d \sigma}{d \chi} = \tfrac{d \sigma}{d \hat{t}} \tfrac{d \hat{t}}{d \chi} = \tfrac{d \sigma}{d \hat{t}} \tfrac{\hat{t}^2}{\hat{s}}$ --- and inserting it in to equation \eqref{eq:analysis:differentialxsec} one obtains 
\begin{empheq}[box=\eqbox]{equation} \label{eq:analysis:differentialxsecdchi}
	\frac{d \sigma}{d \chi} = \int_{\hat{s}_\textrm{min}}^{\hat{s}_\textrm{max}} d \hat{s} \int_{Y_\textrm{min}}^{Y_\textrm{max}} d Y \; x_1 f_1(x_1) \, x_2 f_2(x_2) \frac{d \sigma}{d \hat{t}} \frac{\hat{t}^2}{\hat{s}^2} .
\end{empheq}
The most recent CMS angular analysis \cite{Chatrchyan:2012bf} sets the integration limits to $\left| Y \right| < 1.1$. Then the data is binned in $\hat{s}$, where the most significant bin in the CMS analysis is $\hat{s} > 3 \; \textrm{TeV}$. This analysis then looks for differences between QCD and the effective operator in the $\tfrac{1}{\sigma} \tfrac{d\sigma}{d\chi}$ distribution. These distributions are shown in figure \ref{fig:analysis:dsigmdchi} for QCD, the toy models and their corresponding effective theories.

\subsection{\texorpdfstring{$F_\chi$}{Fchi} Variable}
\label{sec:analysis:fchivariable}
\begin{figure}[!ht]
	\centering
	\includegraphics[width=0.46\textwidth]{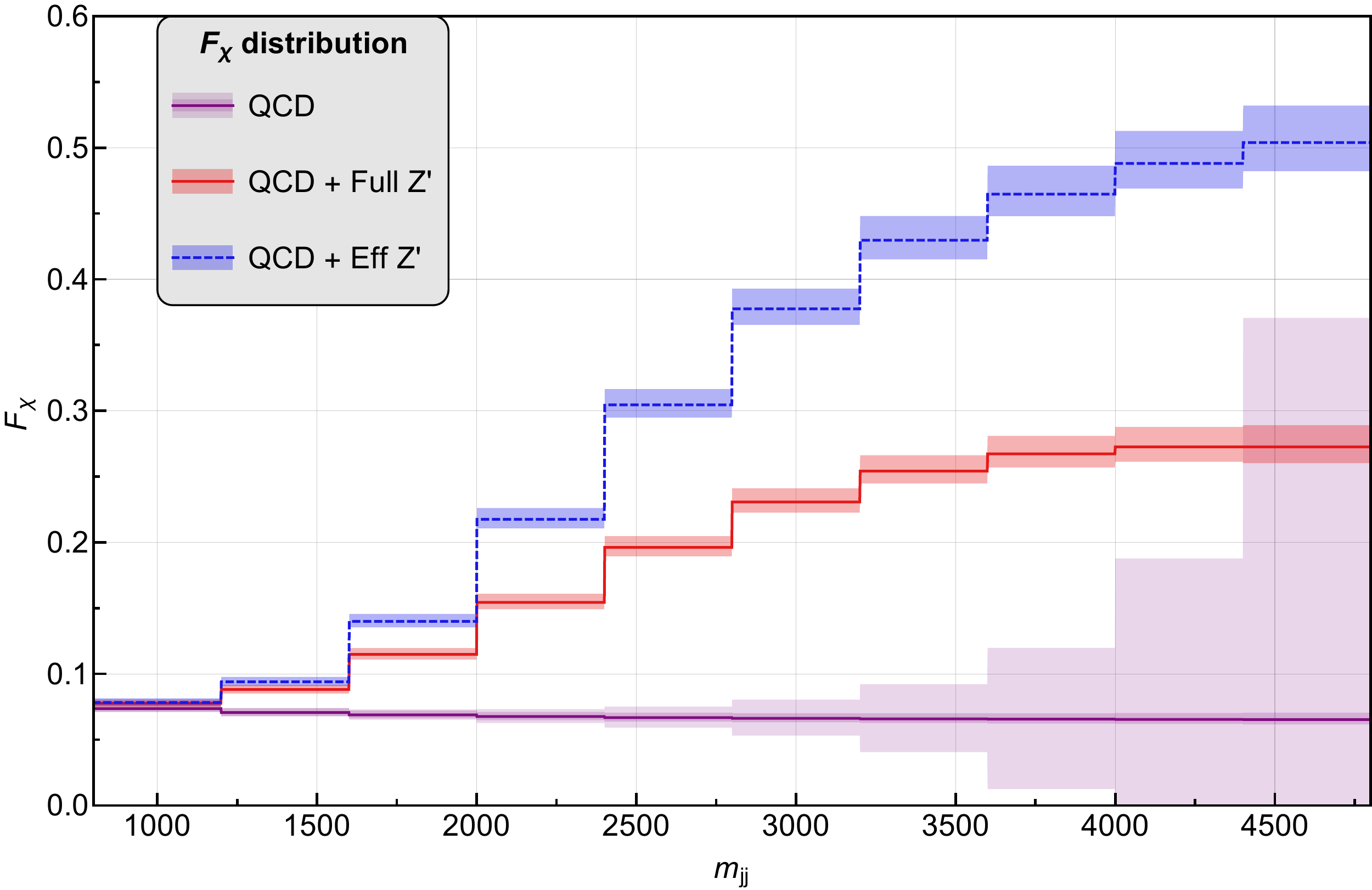}  \hspace{2mm}  
	\includegraphics[width=0.46\textwidth]{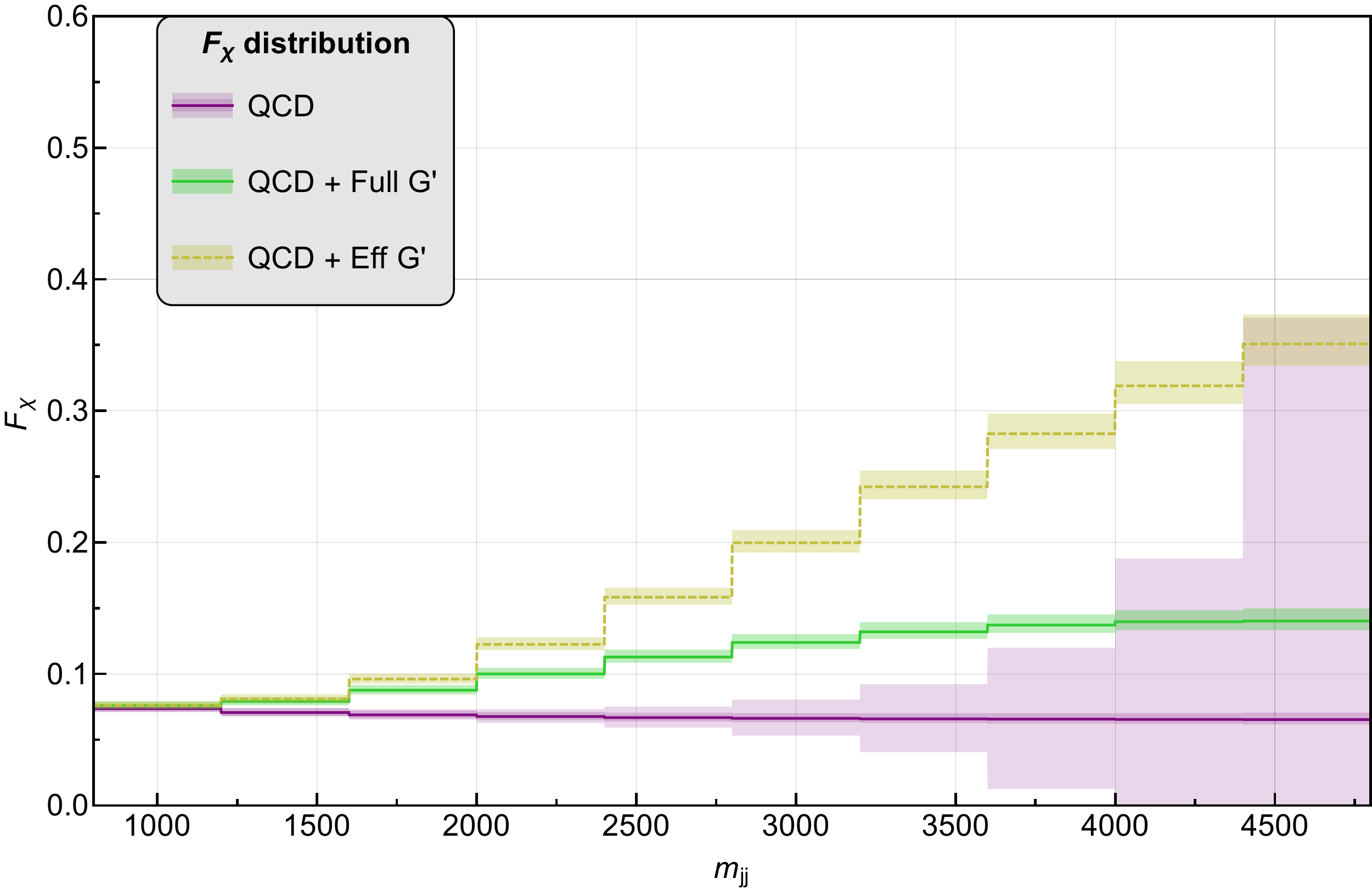} 
	\caption{Reconstruction of the experimental $F_\chi$ distributions for QCD, for the $Z'$ model (left) and the $G'$ model (right) with parameters $m_{Z'} = m_{G'} = 2 \; \textrm{TeV}$ and $g_{Z'} = g_{G'} = \tfrac{\pi}{2}$. The bands around the different distributions represent for QCD the theory error (inner band) and statistical error (outer band). For the full and effective theory the bands represent the theory error, for which more details are given in section \ref{sec:analysis:erroranalysis}.}
	\label{fig:analysis:fchi}
\end{figure}

The ATLAS analyses \cite{Collaboration:2010eza,Aad:2011aj,ATLAS:2012pu} use a single parameter which measures the isotropy of the dijet events. This is defined as 
\begin{equation} \label{eq:analysis:fchidefinition}
	F_\chi \equiv \frac{N_\textrm{central}}{N_\textrm{total}} ,
\end{equation}
where $N_\textrm{central}$ is the number of events in the central region with $1 < \chi < \chi_\textrm{central}$ and $N_\textrm{total}$ is the total number of events with $1 < \chi < \chi_\textrm{max}$. This parameter can depend on $\hat{s}$, for that purpose we explicitly write the $\hat{s}_\textrm{min}$ and $\hat{s}_\textrm{max}$ in equation \eqref{eq:analysis:differentialxsecdchi} and define the integral over $\chi$ as 
\begin{equation} \label{eq:analysis:totalxsec}
	\sigma \left( \chi_\textrm{int} , \hat{s}_\textrm{min}, \hat{s}_\textrm{max} \right) = \int_1^{\chi_\textrm{int}} d \chi \frac{d \sigma}{d \chi} \left( \hat{s}_\textrm{min}, \hat{s}_\textrm{max} \right) .
\end{equation}
The total cross section thus depends on three integration boundaries, from which we can formally define $F_\chi$ as
\begin{empheq}[box=\eqbox]{equation} \label{eq:analysis:fchi}
	F_\chi \left( \hat{s}_\textrm{min}, \hat{s}_\textrm{max} \right) = \frac{\sigma \left( \chi_\textrm{central} , \hat{s}_\textrm{min}, \hat{s}_\textrm{max} \right)}{\sigma \left( \chi_\textrm{max} , \hat{s}_\textrm{min}, \hat{s}_\textrm{max} \right)} .
\end{empheq}
In the most recent ATLAS analysis \cite{ATLAS:2012pu} the event selection criteria $\left| Y \right| < 1.1$ and $\hat{s} > 800 \; \textrm{GeV}$ are used. The boundaries for the $\chi$ limits are $\chi_\textrm{central} = 3.32$ and $\chi_\textrm{max} = 30.0$, the $F_\chi$ parameter is then binned in the dijet invariant mass $m_{jj} = \sqrt{\hat{s}}$. Example distributions are shown in figure \ref{fig:analysis:fchi} for QCD, the toy models and their corresponding effective theories.

\subsection{Error Analysis}
\label{sec:analysis:erroranalysis}
In the next section the comparison between the background --- pure QCD --- and a possible signal is made. For these types of comparisons a detailed account for the different errors affecting the angular distributions is needed. The distributions discussed in sections \ref{sec:analysis:angulardist} and \ref{sec:analysis:fchivariable} depend on ratios of number of events in certain kinematic regions. The error on the specific variable in either of the considered distributions is obtained by propagating the error on the number of events. In the following we describe the error on the number of events coming from different sources and their effect on the distributions discussed in the two previous sections.

In the theoretical limit setting procedure the data is assumed to equal the background prediction including the total error on the background coming from statistic and systematic uncertainties. For the QCD background we consider statistical errors on the number of events to be Poisson distributed. The systematic errors originate from experimental effects and from theoretical uncertainties. The systematic uncertainties from experimental effects are described in the respective analyses \cite{Chatrchyan:2012bf,ATLAS:2012pu} and range up to $15\%$ for the highest mass bin in the angular distribution. For the $F_\chi$ distribution, which is used in the limit setting in the next section, the experimental systematic uncertainties range up to $50\%$. Theoretical uncertainties are estimated by varying the renormalisation and factorisation scales by half and twice their values and by including parton density uncertainties. When these uncertainties are propagated to the angular variables, this results in errors of at most a few percent for both distributions. 

The limit setting for the signal does not involve any statistical errors and solely depends on the systematic uncertainties from theory calculations. As for the QCD background these uncertainties are estimated by varying the renormalisation and factorisation scales by half and twice their values and by including the parton density uncertainties. We find resulting errors which agree with uncertainties found in next-to-leading order calculations for these processes \cite{Gao:2010bb,Gao:2011ha,Gao:2012qpa}. For the signal, which is for each of the two toy models, the errors are similar to the background and range up to $10\%$ when looking at the angular distributions. These distributions, shown in figures \ref{fig:analysis:dsigmdchi} and \ref{fig:analysis:fchi}, include all the errors discussed in this section based on events with a centre of mass energy of $7 \; \textrm{TeV}$ and using $5 \; \textrm{fb}^{-1}$ of integrated luminosity. Other theoretical errors for the effective description are introduced by renormalisation group running and mixing effects \cite{Englert:2014cva}. However, these effects have been estimated to be of the order of $10\%$ for differential dijet cross sections and due to the fact that we are considering ratios of cross sections these errors can be safely neglected in our discussion.

\section{Results}
\label{sec:results}
The goal of this work is to quantify the difference between the full and effective theory exclusions limits in the mass versus coupling plane analogous to figure \ref{fig:effoperators:efftofull}. For this an experimental measure based on the angular analyses needs to be introduced. We observe that there is a significant deviation between the full and effective description for both the experimental angular distributions $\tfrac{1}{\sigma}\tfrac{d\sigma}{d\chi}$ and $F_\chi$ presented in figures \ref{fig:analysis:dsigmdchi} and \ref{fig:analysis:fchi}. The theoretical measure used in this section is based on the $F_\chi$ distribution, similar to the analysis in reference \cite{ATLAS:2012pu}, since this observable is a ratio of the number of events in different angular regions. For the $F_\chi$ distribution many systematic effects cancel, making it a sensitive probe for deviations from QCD.

\begin{figure}[!ht]
	\centering
	\includegraphics[width=0.6\textwidth]{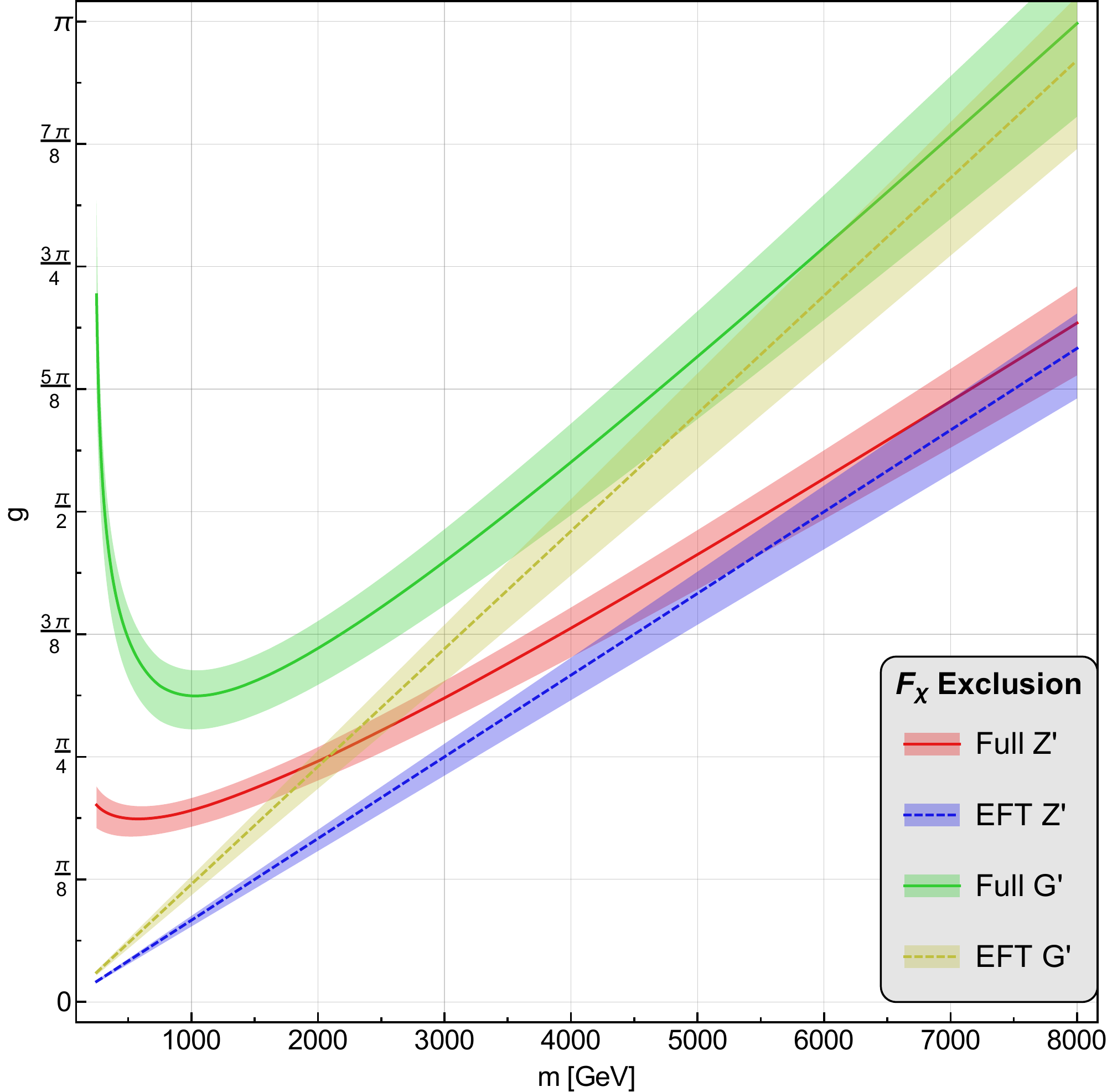} 
	\caption{Comparison of exclusion limits in the mass versus coupling plane between effective theory (dashed lines) and full theory (solid lines). The region above the lines is excluded at $95\%$ confidence level based on a binned $F_\chi$ theoretical measure as described in this section. The bands around the lines show the theory errors on the exclusion regions.}
	\label{fig:results:fchiexclusion}
\end{figure}

In the ATLAS analysis the $F_\chi$ data is binned in the $\hat{s} = m_{jj}^2$ variable as in figure \ref{fig:analysis:fchi} and deviations between experimental data and background predictions are looked for in these bins. The simplest theoretical measure would be taking a single large bin in $\hat{s}$ and performing a $\chi^2$ analysis on difference between the theory predictions for the full and effective theory and the data, see for example reference \cite{Domenech:2012ai}. However, this implies less sensitivity to the kinematic details of the distribution and moreover less similarity with the actual experimental method. Therefore we adopt a more detailed $\chi^2$ measure based on the full set of bins\footnote{The ATLAS analysis uses a different statistical method to look for deviations, namely the tail hunter method \cite{Choudalakis:2011qn}. However, the deviation between this method and a $\chi^2$ analysis is not expected to be significant.}. From figure \ref{fig:analysis:fchi} a reasonable binning is determined to be ranging from $1200 \; \textrm{GeV}$ to $4000 \; \textrm{GeV}$ in $\sqrt{\hat{s}}$ with steps of $400 \; \textrm{GeV}$. Then a $\chi^2$ analysis on the $F_\chi$ variable with the errors as described in section \ref{sec:analysis:erroranalysis} is repeated for different values of the coupling and mass of the toy model. These results are then transformed into a $95\%$ confidence level exclusion contour in the mass versus coupling plane, presented in figure \ref{fig:results:fchiexclusion}.

The results in this section are obtained for a centre of mass energy of $7 \; \textrm{TeV}$ and an integrated luminosity of $5 \; \textrm{fb}^{-1}$. This corresponds roughly to the analysis presented in reference \cite{ATLAS:2012pu} and therefore allows for a good comparison with limits obtained in there. Even though the operators corresponding to the $Z'$ and $G'$ models from equation \eqref{eq:toymodel:efflagrangian} are different from the ones studied in reference \cite{ATLAS:2012pu}, a rough comparison can be made. The limits for the effective description in figure \ref{fig:results:fchiexclusion} correspond to the limits on the effective operator coefficients $\Lambda$. These equal  
\begin{equation} \label{eq:results:oplimit}
	\Lambda_{Z'} = 13.5^{+1.1}_{-0.7} \; \textrm{TeV} , \qquad \qquad \Lambda_{G'} = 9.4^{+1.0}_{-0.6} \; \textrm{TeV} ,
\end{equation}
and we observe an approximate agreement with the results from the ATLAS analysis when correcting for the different definitions used for the four-quark effective operators. In the near future the LHC enters the second phase with a $14 \; \textrm{TeV}$ centre of mass energy for which this analysis is relevant as well. The results for LHC14 are provided in appendix \ref{sec:resultslhc14}, where further details can be found. In the next section the deviation between the full and effective theory is quantified and compared to the effective field theory expansion.

\subsection{EFT Expansion Check}
\label{sec:results:eftexpansion}
In this section the error made by using the effective description for excluding the full theory parameter space is quantified. From a theoretical viewpoint, the error introduced by the effective expansion is governed by the ratio of the transfer energy and the mass of the particle being integrated out as presented in equation \eqref{eq:effexpansion:expansion}. As the series is truncated after the first term, the deviation of the effective partonic cross section compared with the full cross section is expected to be given roughly by $q^2 / m^2$. However, for the limit setting the difference in the total cross sections is also influenced by the parton density functions, the kinematic requirements and the statistical analysis being used. Therefore, the scaling of the deviation in the exclusion limits of the full theory's parameter space is expected to be modified by these effects.

\begin{figure}[!ht]
	\centering
	\includegraphics[width=0.6\textwidth]{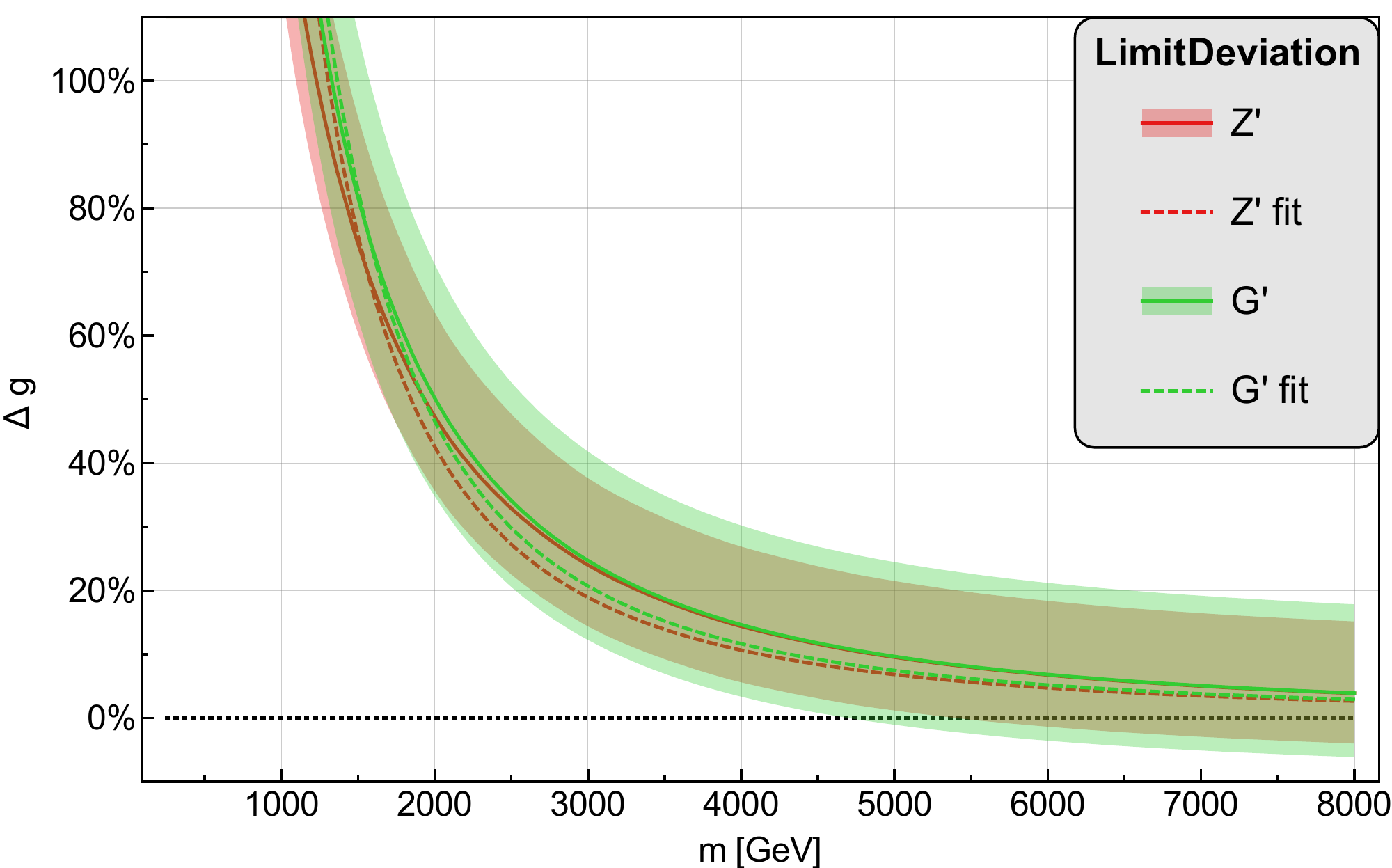} 
	\caption{Deviation of the full theory description with respect to the effective theory for the $F_\chi$ based exclusion. The solid lines show the difference between the full and effective description for the limits on the coupling constants given in equation \eqref{eq:results:couplingdeviation} as a function of the mass. The dashed lines show the fitted function in equation \eqref{eq:results:deviationfit} using the fitted result for $C$. The bands around the solid lines show the theory errors for the deviation.}
	\label{fig:results:deviation}
\end{figure}

\paragraph{Deviation} \mbox{} \\
An interesting quantity to measure is the deviation between the effective and full description is the difference between exclusions limits for the coupling constant for a given mass of the full theory particle. This deviation can be defined as 
\begin{empheq}[box=\eqbox]{equation} \label{eq:results:couplingdeviation}
	\Delta g \equiv \frac{g_\textrm{full} - g_\textrm{eff}}{g_\textrm{eff}} ,
\end{empheq}
and is represented in figure \ref{fig:results:deviation} by the solid lines. The figure shows the deviation for the $F_\chi$ based exclusion described in the previous section and presented as in figure \ref{fig:results:fchiexclusion}. From the interesting observation that the deviation scales to good approximation as $1/m^2$, it is conjectured that $\Delta g$ can be fitted to the function 
\begin{equation} \label{eq:results:deviationfit}
	\Delta g \simeq \frac{C^2}{m^2} .
\end{equation}
This function with a single free parameter $C$ is then fitted to the actual $\Delta g$ in figure \ref{fig:results:deviation} and is represented by the dashed lines. For the $F_\chi$ based exclusions the free parameter equals 
\begin{equation} \label{eq:results:cfitfchi}
	C_{Z'} = 1.31^{+0.20}_{-0.20} \; \textrm{TeV} , \qquad C_{G'} = 1.37^{+0.25}_{-0.21} \; \textrm{TeV} .
\end{equation}
The difference between the $Z'$ and $G'$ models is small, which might indicate that the coefficient $C$ is indeed mainly determined by the effects of the parton densities, the kinematics in the analysis and the statistical method.

\paragraph{Average Transfer Energy} \mbox{} \\
In section \ref{sec:effoperators:eftexpansion} the expansion around the energy transfer was introduced to estimate the validity of the EFT at parton level. In order to gain more insight in the deviation of the effective expansion, an estimate for the average energy transfer in the events considered in the analysis is needed. These averages depend on the kinematic requirements of the angular analyses discussed in the previous section and we present the average values for all the hatted Mandelstam variables. The expressions read
\begin{empheq}[box=\eqbox]{align} \label{eq:analysis:averagemandelstam}
	\left\langle \hat{s} \right\rangle = & \frac{1}{\sigma_\textrm{tot}} \int_{\hat{s}_\textrm{min}}^{\hat{s}_\textrm{max}} d \hat{s} \int_{\chi_\textrm{min}}^{\chi_\textrm{max}} d \chi \, \hat{s} \frac{d^2 \sigma}{d \hat{s} d \chi} \nonumber \\
	\left\langle \hat{t} \right\rangle = & \frac{1}{\sigma_\textrm{tot}} \int_{\hat{s}_\textrm{min}}^{\hat{s}_\textrm{max}} d \hat{s} \int_{\chi_\textrm{min}}^{\chi_\textrm{max}} d \chi \, \frac{- \hat{s}}{1 + \chi} \frac{d^2 \sigma}{d \hat{s} d \chi} \nonumber \\
	\left\langle \hat{u} \right\rangle = & \frac{1}{\sigma_\textrm{tot}} \int_{\hat{s}_\textrm{min}}^{\hat{s}_\textrm{max}} d \hat{s} \int_{\chi_\textrm{min}}^{\chi_\textrm{max}} d \chi \, \frac{- \hat{s} \, \chi}{1 + \chi} \frac{d^2 \sigma}{d \hat{s} d \chi} ,
\end{empheq}
with the normalisation factor $\sigma_\textrm{tot}$ defined as
\begin{equation} \label{eq:results:totalxsex}
	\sigma_\textrm{tot} = \int_{\hat{s}_\textrm{min}}^{\hat{s}_\textrm{max}} d \hat{s} \int_{\chi_\textrm{min}}^{\chi_\textrm{max}} d \chi \frac{d^2 \sigma}{d \hat{s} d \chi} .
\end{equation}
The theoretical averages for an experimental analysis can then be obtained by integrating over the ranges given in the analysis for $\hat{s}$ and $\chi$. For the $F_\chi$ based theoretical measure the results are depicted in table \ref{tab:results:transferenergies} for both the total and the central region in $\chi$.

\begin{table}[!ht]
	\centering
	\begin{tabular}{| c | c | c | c c | c c |}
		\toprule[1pt]
		region & $\sqrt{\left| \left\langle q^2 \right\rangle \right|}$ & QCD & Full $Z'$ & Eff $Z'$ & Full $G'$ & Eff $G'$ \\
		\midrule[1pt]
		\multirow{3}{*}{total} & $\sqrt{\left| \left\langle \hat{s} \right\rangle \right|}$ & $1.43^{+0.16}_{-0.13}$ & $1.45^{+0.16}_{-0.13}$ & $1.47^{+0.16}_{-0.13}$ & $1.44^{+0.16}_{-0.13}$ & $1.45^{+0.16}_{-0.13}$ \\
		& $\sqrt{\left| \left\langle \hat{\scalebox{0.88}{$t$}} \right\rangle \right|}$ & $0.43^{+0.05}_{-0.04}$ & $0.46^{+0.05}_{-0.04}$ & $0.49^{+0.05}_{-0.04}$ & $0.45^{+0.05}_{-0.04}$ & $0.45^{+0.05}_{-0.04}$ \\
		& $\sqrt{\left| \left\langle \hat{u} \right\rangle \right|}$ & $1.36^{+0.15}_{-0.13}$ & $1.37^{+0.15}_{-0.12}$ & $1.38^{+0.15}_{-0.12}$ & $1.37^{+0.15}_{-0.13}$ & $1.37^{+0.15}_{-0.13}$ \\
		\midrule[1pt]
		\multirow{3}{*}{central} & $\sqrt{\left| \left\langle \hat{s} \right\rangle \right|}$ & $1.43^{+0.16}_{-0.14}$ & $1.50^{+0.15}_{-0.12}$ & $1.58^{+0.15}_{-0.12}$ & $1.46^{+0.16}_{-0.13}$ & $1.49^{+0.16}_{-0.13}$ \\
		& $\sqrt{\left| \left\langle \hat{\scalebox{0.88}{$t$}} \right\rangle \right|}$ & $0.82^{+0.10}_{-0.08}$ & $0.88^{+0.09}_{-0.07}$ & $0.93^{+0.09}_{-0.07}$ & $0.85^{+0.09}_{-0.07}$ & $0.87^{+0.09}_{-0.08}$ \\
		& $\sqrt{\left| \left\langle \hat{u} \right\rangle \right|}$ & $1.17^{+0.13}_{-0.11}$ & $1.22^{+0.12}_{-0.10}$ & $1.28^{+0.12}_{-0.10}$ & $1.19^{+0.13}_{-0.11}$ & $1.21^{+0.13}_{-0.11}$ \\
		\bottomrule[1pt]
	\end{tabular}
	\caption{Average transfer energies in TeV, which are the hatted Mandelstam variables for both the total region with $\chi < \chi_\textrm{max}$ and the central region with $\chi < \chi_\textrm{central}$. Presented are the absolute values of the Mandelstam variables, however, they are constrained to $\hat{s} > 0$ and $\hat{t}, \hat{u} < 0$. The values correspond to the model parameters $m_{Z'} = m_{G'} = 2 \; \textrm{TeV}$ and $g_{Z'} = g_{G'} = \tfrac{\pi}{2}$. The errors on these average transfer energies are the theory uncertainties from scale variation and parton density uncertainties.}
	\label{tab:results:transferenergies}
\end{table}

The experimental collaborations can determine each of the average transfer energies by using the kinematic information on an event by event basis. For each event the kinematic variables $\hat{s}$ and $\chi$ are known from measurements on the dijet system. Then, with the use of equation \eqref{eq:analysis:chi} and the sum of Mandelstam variables $\hat{s} + \hat{t} + \hat{u} = 0$ (for vanishing quark masses) the quantities $\hat{s}$, $\hat{t}$ and $\hat{u}$ can be determined for each event. Finally, averaging over all events leads to the determination of the average transfer energies in the experiment.

The detailed knowledge of the average transfer energies in the events allows for the observation that the concept of a sharp cut-off of the effective theory expansion at $\left \langle q^2 \right \rangle = m^2$ is not a sensible approach. Even for masses above this cut-off, where the effective description is generally considered valid \cite{Englert:2014cva,Dreiner:2013vla,Domenech:2012ai,Bessaa:2014jya}, from figures \ref{fig:results:fchiexclusion} and \ref{fig:results:deviation} we observe significant deviations in the bound on the coupling constant. Compared to the usual theoretical errors, which arise from QCD effects and parton density uncertainties, the errors made by applying the effective description dominate up to a mass of roughly $5 \; \textrm{TeV}$. This exactly corresponds to the mass range of interesting BSM models in the light of LHC phenomenology. Therefore, it is suggested that errors from employing the effective operator approach are presented in combination with the resulting bounds. Alternatively, for special classes of BSM states a rescaling procedure to recast the existing experimental effective operator limits is described in the next section.

\subsection{Limit Recast}
\label{sec:results:limitrecast}
The ultimate goal of the quantification of the deviation is to re-analyse existing limits from effective operators. The full theory effects could be included by performing a rescaling based on the fitted result for the deviation in equation \eqref{eq:results:deviationfit}. These results have been obtained for $Z'$ and $G'$ toy models and could be applied to similar models as well. This would then lead to a more reliable exclusion limit in the mass versus coupling plane for these models, which is crucial for scrutinising a model's parameter space. In the previous section the deviation was quantified for the ATLAS analysis in reference \cite{ATLAS:2012pu}. In appendix \ref{sec:recastexample} we perform a recast for a colour octet considered in \cite{Redi:2013eaa}, which has been excluded using these ATLAS limits on four-quark effective operators. This example illustrates how considering the full theory effects leads to more reliable exclusion limits.

Consequently, experiments are urged to apply their angular dijet analyses to full models --- like the $Z'$ and $G'$ which cover a large class of BSM models --- as well. Thereby allowing for a trustworthy quantification of the deviation between the full and effective description, which can then be applied to similar BSM models.

\section{Conclusions}
\label{sec:conclusion}
The interpretation of effective operator limits from hadron colliders for beyond the Standard Model theories with energy scales around the TeV scale is a delicate matter. It is argued that effective operator limits for BSM theories at and around the TeV scale should be more carefully interpreted in the mass versus coupling plane. In this work the pitfalls are identified and methods to reliably interpret the experimental limits are described.

For a correct interpretation of effective operator limits at the Large Hadron Collider it is necessary that experimental collaborations provide information on the average transfer energies in their analyses. In section \ref{sec:results:eftexpansion} a procedure which achieves this is outlined for the existing angular analyses. Furthermore, the collaborations are invited to apply the limit setting procedure to generic models like the $Z'$ and $G'$ as well. This allows for more accurate limit setting in the mass versus coupling plane for specific BSM models with similarities to these models. From these results it is deduced that the concept of a sharp mass threshold above which the effective description is valid is not a sensible approach, rather a continuous deviation from the effective limit is observed. The error introduced by this deviation dominates over the standard errors from QCD corrections and parton density uncertainties for masses of BSM particles up to $5 \; \textrm{TeV}$. Therefore, it is advisable to provide a detailed account of the errors which arise in effective descriptions at hadron colliders.

As an alternative, the effective operator limits can be correctly interpreted in the full theory through a rescaling procedure based on the quantified deviation in section \ref{sec:results:eftexpansion}. This deviation scales as the inverse square of the mass of the BSM particle considered, and can be fitted using the experimental analyses for toy models. The observation that the deviation is not very model dependent implies that this rescaling procedure can be used for a plethora of BSM particles within a reasonable accuracy. In conclusion, a method has been presented which allows for a more reliable scrutinising of BSM parameter spaces while using four-quark effective operator limits.

\acknowledgments
The author of this paper is grateful for discussions with Caterina Doglioni and Marco Tonini. The author has been partially supported by the Deutsche Forschungsgemeinschaft within the Collaborative Research Centre SFB 676 "Particles, Strings, Early Universe". This work and its results are licensed under the Creative Commons Attribution 4.0 International License \cite{website:cclicense}.

\newpage
\appendix

\section{Toy Model}
\label{sec:toymodel}
The toy model should be as simple as possible while still reproducing the interesting parts of realistic BSM models. Based on $Z'$ models we can construct a single boson which couples uniformly to quarks, also known as the hadronic $Z'$. Equivalently also a partner for the gluon can be constructed, denoted as $G'$. These spin-one bosons will be used and their interactions with the Standard Model quarks are governed by the  Lagrangians
\begin{align} \label{eq:toymodel:lagrangian}
	\mathcal{L}_{Z'} & \subset - m_{Z'}^2 Z'^\mu Z'_\mu + g_{Z'} \, \bar{q}_i \gamma^\mu \delta_{ij} q_j \, Z'_\mu \nonumber \\
	\mathcal{L}_{G'} & \subset - m_{G'}^2 G'^{a\mu} G'^a_\mu + g_{G'} \, \bar{q}_i \gamma^\mu T^a_{ij} q_j \, G'^a_\mu .
\end{align}
In here $g_{Z'}$ and $g_{G'}$ are the coupling constants and $i,j$ the colour indices. For these toy models the different transformations under gauge groups and the charges of the quarks are not taken into account. This is not relevant for the analysis in this work, for a comprehensive description discussing anomalies see \cite{Dobrescu:2013cmh}. Another relevant property of these particles are their masses which are denoted as $m_{Z'}$ and $m_{G'}$ respectively. Together with the coupling constants they form the fundamental parameters of this toy model. The Feynman rule for the $Z'\bar{q}q$ and $G'\bar{q}q$ couplings are
\begin{equation} \label{eq:toymodel:feynmanrules}
	\raisebox{-10mm}{
		\begin{tikzpicture}[line width=1.4 pt, scale=1]
			\draw[fermionbar] (-40:1)--(0,0);
			\draw[fermion] (40:1)--(0,0);
			\draw[vector] (180:1)--(0,0);
			\draw[fill=black] (0,0) circle (1.5pt);
			\node at (40:1.2) {$\bar{q}_i$};
			\node at (-40:1.2) {$q_j$};
			\node at (180:1.2) {$Z'_\mu$};
			\node at (0:2.2) {$= i g_{Z'} \gamma^\mu \delta_{ij}$};
		\end{tikzpicture}
		\qquad \qquad
		\begin{tikzpicture}[line width=1.4 pt, scale=1]
			\draw[fermionbar] (-40:1)--(0,0);
			\draw[fermion] (40:1)--(0,0);
			\draw[gluon] (180:1)--(0,0);
			\draw[fill=black] (0,0) circle (1.5pt);
			\node at (40:1.2) {$\bar{q}_i$};
			\node at (-40:1.2) {$q_j$};
			\node at (180:1.2) {$G'_\mu$};
			\node at (0:2.2) {$= i g_{G'} \gamma^\mu T^a_{ij} .$};
		\end{tikzpicture}
	}
\end{equation}
In the rest of the appendix two additional important properties of the toy model are discussed. First, in the next two sections the width and its effect in the propagator are calculated for both the $Z'$ and the $G'$ and secondly in section \ref{sec:toymodel:operators} the effective operators generated by this toy model are derived. 

\subsection{Widths}
\label{sec:toymodel:width}
A relevant property of any particle in detector based experiments is the width, it influences the detectability in resonance searches. Though, also the width may have an impact on the correctness of the effective description. For our simple bosons the partial widths for decaying into a single $q \bar{q}$ pair are given by
\begin{empheq}[box=\eqbox]{align} \label{eq:toymodel:widths}
	\Gamma_{Z' \to q \bar{q}} & = \alpha_{Z'} \frac{m_{Z'}^2 + 2 m_q^2}{m_{Z'}^2} \sqrt{m_{Z'}^2 - 4 m_q^2} \nonumber \\
	\Gamma_{G' \to q \bar{q}} & = \frac{\alpha_{G'}}{6} \frac{m_{G'}^2 + 2 m_q^2}{m_{G'}^2} \sqrt{m_{G'}^2 - 4 m_q^2} .
\end{empheq}

\paragraph{Calculation} \mbox{} \\
The starting point for calculating the width of the decay $X \to q \bar{q}$ is the equation
\begin{equation} \label{eq:toymodel:widthcalc:start}
	\Gamma_{X \to q \bar{q}} = \frac{1}{8 \pi} \frac{|\vec{p}_{1,2}|}{m_X^2} \int \frac{d \Omega_{cm}}{4 \pi} \left| \mathcal{M}_{X \to q \bar{q}} \right|^2 ,
\end{equation}
where for equal quark masses the relevant kinematic variables --- assuming incoming momentum $k$ and outgoing momenta $p_1$ and $p_2$ --- in this process are
\begin{align} \label{eq:toymodel:widthcalc:kinematic}
	& k^2 = m_X^2 \qquad p_1^2 = p_2^2 = m_q^2 \qquad 2p_1 \cdot p_2 = m_X^2 - 2m_q^2 \nonumber \\
	& 2 k \cdot p_1 = 2 k \cdot p_2 = m_X^2 \qquad |\vec{p}_1| = |\vec{p}_2| = \frac{1}{2} \sqrt{m_X^2 - 4 m_q^2} .
\end{align}
For the $Z'$ decay the amplitude equals
\begin{equation} \label{eq:toymodel:widthcalc:matrix}
	\mathcal{M}_{Z' \to q \bar{q}} = i g_{Z'} \bar{u}_i(p_1) \gamma^\mu \delta_{ij} v_j(p_2) \epsilon_\mu(k) , 
\end{equation}
similarly for the $G'$ decay with the replacements $g_{Z'} \to g_{G'}$, $\delta_{ij} \to T^a_{ij}$ and $\epsilon_\mu(k) \to \epsilon^a_\mu(k)$. Then square the amplitude and average over initial spin and colour to obtain
\begin{equation} \label{eq:toymodel:widthcalc:avgmatrix}
	\left| \overline{\mathcal{M}_{Z' \to q \bar{q}}} \right|^2 = g_{Z'}^2 \left( -g_{\mu\nu} + \frac{k_\mu k_\nu}{m_{Z'}^2} \right) \tr \left[ (\slashed{p}_1 + m_q) \gamma^\mu (\slashed{p}_2 - m_q) \gamma^\nu) \right] .
\end{equation}
The same can be obtained for $G'$ with a different factor due to the colour structure and averaging over initial colour. This leads to the identification $\left| \overline{\mathcal{M}_{G' \to q \bar{q}}} \right|^2 = \tfrac{1}{6} \left| \overline{\mathcal{M}_{Z' \to q \bar{q}}} \right|^2$ with the obvious $Z' \to G'$ replacements. Evaluating the trace
\begin{equation} \label{eq:toymodel:widthcalc:trace}
	\tr \left[ (\slashed{p}_1 + m_q) \gamma^\mu (\slashed{p}_2 - m_q) \gamma^\nu) \right] = 4 \left[ p_1^\mu p_2^\nu + p_1^\nu p_2^\mu - g^{\mu\nu} \left( m_q^2 + p_1 \cdot p_2 \right) \right]
\end{equation}
and using the kinematic expressions from equation \eqref{eq:toymodel:widthcalc:kinematic} reduces the averaged matrix element to
\begin{equation} \label{eq:toymodel:widthcalc:finalmatrix}
	\left| \overline{\mathcal{M}_{Z' \to q \bar{q}}} \right|^2 = 4 g_{Z'}^2 \left[ m_{Z'}^2 + 2 m_q^2 \right] .
\end{equation}
Plugging this expression into equation \eqref{eq:toymodel:widthcalc:start} leads to final result given in equation \eqref{eq:toymodel:widths}.

\subsection{Propagator}
\label{sec:toymodel:propagator} 
For the calculation of the dijet cross sections in appendix \ref{sec:dijetxsec} and for the determination of the effective operator coefficients in the next section a proper definition for the propagator including the width is needed. In general for $Z'$ like models large widths are a possibility and the usual Breit-Wigner propagator using the narrow-width approximation is not valid. Instead we adopt the methods developed in \cite{Gounaris:1968mw,Kuhn:1990ad}, which imply that for the $Z'$ case the propagator equals
\begin{empheq}[box=\eqbox]{equation} \label{eq:toymodel:propagator}
	\Pi_{Z'} \! \left( q^2 \right) = \frac{-i g_{\mu\nu}}{q^2 - m_{Z'}^2 + i \sqrt{q^2} \, \Gamma_{Z'} \! \left( q^2 \right)} .   
\end{empheq}
The choice for the role of the width in the propagator is not unique, which stems from our ignorance about higher order corrections. However, this choice provides a good description for a large range of transfer energies \cite{Kuhn:1990ad}, where the typical Breit-Wigner propagator would break down.

In this expression the width depends on the transferred momentum in the propagator $q^2$, which for the dijet cross sections may equal either $\hat{s}$, $\hat{t}$ or $\hat{u}$. At leading order the width is given by
\begin{equation} \label{eq:toymodel:dependentwidth}
	\Gamma_{Z'} \! \left( q^2 \right) = \sum_i \Gamma_{Z' \to q_i \bar{q}_i} \, \frac{\left( q^2 - 4 m_{q_i}^2 \right)^\frac{3}{2}}{\left( m_{Z'}^2 - 4 m_{q_i}^2 \right)^\frac{3}{2}} \frac{m_{Z'}^2}{q^2} .
\end{equation}
The width $\Gamma_{Z' \to q_i \bar{q}_i}$ is given in equation \eqref{eq:toymodel:widths} in the previous section and the sum is over all six quark flavours. The results for the $G'$ model are exactly the same and are obtained using the replacement $Z' \to G'$. In the rest of the calculations involving the width or the propagator, the quark masses are neglected, which leads to 
\begin{equation} \label{eq:toymodel:dependentwidthnomass}
	\Gamma_{Z'} \! \left( q^2 \right) = 6 \alpha_{Z'} \sqrt{q^2} .
\end{equation}

\subsection{Effective Operators} 
\label{sec:toymodel:operators}
The full theory is given in \eqref{eq:toymodel:lagrangian} and from this we can obtain an effective theory by integrating out the $Z'$ or $G'$ boson. Among other higher-dimensional operators these two are generated
\begin{equation} \label{eq:toymodel:efflagrangian}
 \mathcal{L}^\textrm{eff} = c_{Z'} \left[ \bar{q}_i \gamma^\mu \delta_{ij} q_j \right]^2 + c_{G'} \left[ \bar{q}_i \gamma^\mu T^a_{ij} q_j \right]^2 .
\end{equation}
The Feynman rule for each of the operators reads
\begin{equation} \label{eq:toymodel:efffeynmanrule}
	\raisebox{-10mm}{
		\begin{tikzpicture}[line width=1.4pt, scale=1]
			\draw[fermionbar] (45:1)--(0,0);
			\draw[fermionbar] (-45:1)--(0,0);
			\draw[fermion] (135:1)--(0,0);
			\draw[fermion] (-135:1)--(0,0);
			\draw[fill=black] (0,0) circle (0.9mm);
			\draw[fill=gray] (0,0) circle (1mm);
			\node at (135:1.2) {$q_i$};
			\node at (-135:1.2) {$q_j$};
			\node at (45:1.2) {$q_k$};
			\node at (-45:1.2) {$q_l$};
			\node at (0:2.9) {$= 2 \, i \, c_{Z'} \, \gamma^\mu \, \delta_{ik} \, \gamma_\mu \, \delta_{jl} ,$};
		\end{tikzpicture}
	}
\end{equation}
where for the $G'$ boson $\delta_{ij}$ is replaced by $T^a_{ij}$. Note that the combination where $k$ and $l$ are interchanged also exists. From the calculation below when matching the full theory onto this effective theory we find that the coefficients equal
\begin{empheq}[box=\eqbox]{equation} \label{eq:toymodel:effcoefficients}
	c_{Z'} = - \frac{g_{Z'}^2}{2 m_{Z'}^2}, \qquad c_{G'} = - \frac{g_{G'}^2}{2 m_{G'}^2} .
\end{empheq}
It is important to note here that the effective operator coefficient does not depend on the width of the $Z'$ or $G'$ particle. The width only enters at non-leading order in the effective expansion of the transfer energy over the mass of the $Z'$ or $G'$ particle.

\paragraph{Calculation} \mbox{} \\
The starting point for the matching are equation \eqref{eq:toymodel:lagrangian} for the $Z'$ and $G'$ bosons and equation \eqref{eq:toymodel:efflagrangian} for the effective theory. For the matching procedure the process $q_i q_j \to q_i q_j$ is used, this only leaves the $t$-channel diagram and simplifies the calculation. In the full theory we have for this amplitude in the case of the $Z'$ 
\begin{equation} \label{eq:toymodel:effcalc:fullamp}
	\mathcal{M}_{ij \to ij}^\textrm{full} = \bar{u}_k(k_3) \left[ i g_{Z'} \gamma^\mu \delta_{ki} \right] u_i(k_1) \frac{-i g_{\mu\nu}}{q^2 - m_{Z'}^2 + i \sqrt{q^2} \, \Gamma_{Z'} \! \left( q^2 \right)} \bar{u}_l(k_4) \left[ i g_{Z'} \gamma^\nu \delta_{kj} \right] u_j(k_2) .
\end{equation}
In the effective theory we find --- using the Feynman rule from equation \eqref{eq:toymodel:efffeynmanrule} --- the amplitude
\begin{equation} \label{eq:toymodel:effcalc:effamp}
	\mathcal{M}_{ij \to ij}^\textrm{eff} = 2 i c_{Z'} \bar{u}_k(k_3) \left[ \gamma^\mu \delta_{ki} \right] u_i(k_1) \bar{u}_l(k_4) \left[ \gamma_\mu \delta_{lj} \right] u_j(k_2) .
\end{equation}
Expanding the propagator around $q^2 = 0$ in the full theory gives
\begin{equation} \label{eq:toymodel:effcalc:propexp}
	\frac{1}{q^2 - m_{Z'}^2 + i \sqrt{q^2} \, \Gamma_{Z'} \! \left( q^2 \right)} = - \frac{1}{m_{Z'}^2} \left[ 1 + \frac{q^2}{m_{Z'}^2} \left( 1 + i \frac{\Gamma_{Z'}}{m_{Z'}} \right) + \cdots \right] .
\end{equation}
Then taking the leading order term from this equation leads to the matched coefficients in equation \eqref{eq:toymodel:effcoefficients}. The calculation for $G'$ follows exactly the same procedure, however, with the replacements $Z' \to G'$ and $\delta_{ij} \to T^a_{ij}$.

\section{Dijet Cross Sections}
\label{sec:dijetxsec}
In this appendix the partonic cross sections for dijet production at the LHC are calculated and tabulated for QCD in combination with the toy model from appendix \ref{sec:toymodel}. Knowing the exact and analytical expressions for all these cross section is essential for the understanding of the experimental limits and the transition between effective and full theory. Since the toy model involves only quarks as external particles for the dijet production, interference with QCD amplitudes involving external gluons is not present. Therefore these processes are presented first and can be directly obtained from the literature \cite{Eichten:1984eu,Domenech:2012ai,Davidson:2013fxa}, the analytic cross sections differential in $\hat{t}$ are
\begin{empheq}[box=\eqbox]{align} \label{eq:dijetxsec:gluons}
	\frac{d \sigma}{d \hat{t}} (g q_i \to g q_i)_\textrm{QCD} & = \frac{4 \pi \alpha_s^2}{9 \hat{s}^2} \left[ - \frac{\hat{u}}{\hat{s}} - \frac{\hat{s}}{\hat{u}} + \frac{9}{4} \frac{\hat{s}^2 + \hat{u}^2}{\hat{t}^2} \right] \nonumber \\
	\frac{d \sigma}{d \hat{t}} (g g \to q_i \bar{q}_i)_\textrm{QCD} & = \frac{\pi \alpha_s^2}{6 \hat{s}^2} \left[ \frac{\hat{u}}{\hat{t}} + \frac{\hat{t}}{\hat{u}} - \frac{9}{4} \frac{\hat{t}^2 + \hat{u}^2}{\hat{s}^2} \right] \nonumber \\
	\frac{d \sigma}{d \hat{t}} (g g \to g g)_\textrm{QCD} & = \frac{9 \pi \alpha_s^2}{2 \hat{s}^2} \left[ 3 - \frac{\hat{t} \hat{u}}{\hat{s}^2} - \frac{\hat{s} \hat{u}}{\hat{t}^2} - \frac{\hat{s} \hat{t}}{\hat{u}^2} \right] .
\end{empheq}
In this work all partonic cross sections will be presented differential in $\hat{t}$, because of their simple structure and easy convolution with the parton density functions in the performed analysis. The relevant production processes only involving external quarks are $q_i q_i \to q_i q_i$ and $q_i q_j \to q_i q_j$ where $i \neq j$. These also include interference effects between QCD and the toy model and therefore need a dedicated calculation. The full details of the calculation are not presented, but a rigorous outline is given in the paragraphs below. At the end of this appendix in equation \eqref{eq:dijetxsec:results} the resulting cross sections are presented. 

\paragraph{Amplitudes} \mbox{} \\
Now we discuss the production processes $q_i q_i \to q_i q_i$ and $q_i q_j \to q_i q_j$ where $i \neq j$. The first takes place through $t$- and $u$-channel exchange, whereas the second is an exact copy of the first with only $t$-channel exchange. Hence the calculation is done only for the first process and for the second process the contributions from $t$-channel exchange are then extracted. As a starting point, all amplitudes relevant for the process are listed for QCD, the full theory and the effective theory (both $t$-channel and $u$-channel)
\begin{align} \label{eq:dijetxsec:amplitudes}
	\mathcal{M}_\textrm{QCD}^{\hat{t}} & = i \frac{g_s^2}{\hat{t}} \left[ \bar{u}_i(k_3) \gamma^\mu T^a_{ij} u_j(k_1) \right] \left[ \bar{u}_k(k_4) \gamma_\mu T^a_{kl} u_l(k_2) \right] \nonumber \\
	\mathcal{M}_\textrm{QCD}^{\hat{u}} & = - i \frac{g_s^2}{\hat{u}} \left[ \bar{u}_i(k_4) \gamma^\mu T^a_{ij} u_j(k_1) \right] \left[ \bar{u}_k(k_3) \gamma_\mu T^a_{kl} u_l(k_2) \right] \nonumber \\
	\mathcal{M}_\textrm{full}^{\hat{t}} & = i \frac{g_{Z'}^2}{\hat{t} - m_{Z'}^2 + i \sqrt{\hat{t}} \, \Gamma_{Z'} \! \left( \hat{t} \right)} \left[ \bar{u}_i(k_3) \gamma^\mu \delta_{ij} u_j(k_1) \right] \left[ \bar{u}_k(k_4) \gamma_\mu \delta_{kl} u_l(k_2) \right] \nonumber \\
	\mathcal{M}_\textrm{full}^{\hat{u}} & = - i \frac{g_{Z'}^2}{\hat{u} - m_{Z'}^2 + i \sqrt{\hat{u}} \, \Gamma_{Z'} \! \left( \hat{u} \right)} \left[ \bar{u}_i(k_4) \gamma^\mu \delta_{ij} u_j(k_1) \right] \left[ \bar{u}_k(k_3) \gamma_\mu \delta_{kl} u_l(k_2) \right] \nonumber \\
	\mathcal{M}_\textrm{eff}^{\hat{t}} & = 2 i c_{Z'} \left[ \bar{u}_i(k_3) \gamma^\mu \delta_{ij} u_j(k_1) \right] \left[ \bar{u}_k(k_4) \gamma_\mu \delta_{kl} u_l(k_2) \right] \nonumber \\
	\mathcal{M}_\textrm{eff}^{\hat{u}} & = - 2 i c_{Z'} \left[ \bar{u}_i(k_4) \gamma^\mu \delta_{ij} u_j(k_1) \right] \left[ \bar{u}_k(k_3) \gamma_\mu \delta_{kl} u_l(k_2) \right] .
\end{align}
For the coloured resonance $G'$ one needs to make the replacements $Z' \to G'$ and $\delta_{ij} \to T^a_{ij}$ in the last four amplitudes. The different colour structure affects the interference terms and some of those may be non-zero for the $G'$ where they would vanish for the $Z'$. We allow the effective operator coefficients $c_{Z'}$ and $c_{G'}$ from equation \eqref{eq:toymodel:effcoefficients} to be complex, furthermore the full theory propagators also include imaginary parts proportional to the width.

\paragraph{Definitions} \mbox{} \\
Per process we want to calculate the spin and colour averaged amplitude
\begin{equation} \label{eq:dijetxsec:summatrix}
	\left| \overline{\mathcal{M}} \right|^2 = \frac{1}{3^2} \sum_\textrm{colour} \frac{1}{2^2} \sum_\textrm{spin} \mathcal{M}_X \mathcal{M}_Y^* ,
\end{equation}
where $\mathcal{M}_X$ and $\mathcal{M}_Y$ are a combination of any of the amplitudes from equations \eqref{eq:dijetxsec:amplitudes}.

Some useful traces, where $k_1$ and $k_2$ are incoming momenta and $k_3$ and $k_4$ are outgoing momenta, are given by
\begin{align} \label{eq:dijetxsec:gammatraces}
	\textrm{tr} \left[ \slashed{k}_3 \gamma^\mu \slashed{k}_1 \gamma^\nu \right] \cdot \textrm{tr} \left[ \slashed{k}_4 \gamma_\mu \slashed{k}_2 \gamma_\nu \right] & = 8 \left( \hat{s}^2 + \hat{u}^2 \right) \nonumber \\
	\textrm{tr} \left[ \slashed{k}_4 \gamma^\mu \slashed{k}_1 \gamma^\nu \right] \cdot \textrm{tr} \left[ \slashed{k}_3 \gamma_\mu \slashed{k}_2 \gamma_\nu \right] & = 8 \left( \hat{s}^2 + \hat{t}^2 \right) \nonumber \\
	\textrm{tr} \left[ \slashed{k}_3 \gamma^\mu \slashed{k}_1 \gamma^\nu \slashed{k}_4 \gamma_\mu \slashed{k}_2 \gamma_\nu \right] & = - 8 \hat{s}^2 .
\end{align}
Moreover, for this momenta configuration and all initial and final state particles massless we have the differential cross section
\begin{equation} \label{eq:dijetxsec:xsec}
	\frac{d \sigma}{d \hat{t}} = \frac{\left| \overline{\mathcal{M}} \right|^2}{16 \pi \hat{s}^2} .
\end{equation}

\paragraph{Squared Amplitudes} \mbox{} \\
The calculation of squaring the amplitudes from equation \eqref{eq:dijetxsec:amplitudes} can be split up in a pre-factor and four spinor structures ($t$-channel colour octet, $u$-channel colour octet, $t$-channel colour singlet and $u$-channel colour singlet)
\begin{align} \label{eq:dijetxsec:spinors}
	\mathcal{M}_{(8)}^{\hat{t}} & = \left[ \bar{u}_i(k_3) \gamma^\mu T^a_{ij} u_j(k_1) \right] \left[ \bar{u}_k(k_4) \gamma_\mu T^a_{kl} u_l(k_2) \right] \nonumber \\
	\mathcal{M}_{(8)}^{\hat{u}} & = \left[ \bar{u}_i(k_4) \gamma^\mu T^a_{ij} u_j(k_1) \right] \left[ \bar{u}_k(k_3) \gamma_\mu T^a_{kl} u_l(k_2) \right] \nonumber \\
	\mathcal{M}_{(1)}^{\hat{t}} & = \left[ \bar{u}_i(k_3) \gamma^\mu \delta_{ij} u_j(k_1) \right] \left[ \bar{u}_k(k_4) \gamma_\mu \delta_{kl} u_l(k_2) \right] \nonumber \\
	\mathcal{M}_{(1)}^{\hat{u}} & = \left[ \bar{u}_i(k_4) \gamma^\mu \delta_{ij} u_j(k_1) \right] \left[ \bar{u}_k(k_3) \gamma_\mu \delta_{kl} u_l(k_2) \right] .
\end{align}
To calculate all contributions from equation \eqref{eq:dijetxsec:amplitudes} to the $q_i q_i \to q_i q_i$ process all possible sixteen contractions from equation \eqref{eq:dijetxsec:summatrix} are needed. These are summarised as
\begin{align} \label{eq:dijetxsec:sqmatrix}
	& \left| \overline{\mathcal{M}_{(8)}^{\hat{t}}} \right|^2 = \frac{4}{9} \left( \hat{s}^2 + \hat{u}^2 \right) & & \left| \overline{\mathcal{M}_{(8)}^{\hat{u}}} \right|^2 = \frac{4}{9} \left( \hat{s}^2 + \hat{t}^2 \right) \nonumber \\
	& \left| \overline{\mathcal{M}_{(1)}^{\hat{t}}} \right|^2 = 2 \left( \hat{s}^2 + \hat{u}^2 \right) & & \left| \overline{\mathcal{M}_{(1)}^{\hat{u}}} \right|^2 = 2 \left( \hat{s}^2 + \hat{t}^2 \right) \nonumber \\
	& \overline{\mathcal{M}_{(8)}^{\hat{t}} {\mathcal{M}_{(8)}^{\hat{u}}}^*} = \overline{\mathcal{M}_{(8)}^{\hat{u}} {\mathcal{M}_{(8)}^{\hat{t}}}^*} = \frac{4}{27} \hat{s}^2 & & \overline{\mathcal{M}_{(1)}^{\hat{t}} {\mathcal{M}_{(1)}^{\hat{u}}}^*} = \overline{\mathcal{M}_{(1)}^{\hat{u}} {\mathcal{M}_{(1)}^{\hat{t}}}^*} = - \frac{2}{3} \hat{s}^2 \nonumber \\
	& \overline{\mathcal{M}_{(8)}^{\hat{t}} {\mathcal{M}_{(1)}^{\hat{t}}}^*} = \overline{\mathcal{M}_{(1)}^{\hat{t}} {\mathcal{M}_{(8)}^{\hat{t}}}^*} = 0 & & \overline{\mathcal{M}_{(8)}^{\hat{u}} {\mathcal{M}_{(1)}^{\hat{u}}}^*} = \overline{\mathcal{M}_{(1)}^{\hat{u}} {\mathcal{M}_{(8)}^{\hat{u}}}^*} = 0 \nonumber \\
	& \overline{\mathcal{M}_{(8)}^{\hat{t}} {\mathcal{M}_{(1)}^{\hat{u}}}^*} = \overline{\mathcal{M}_{(1)}^{\hat{u}} {\mathcal{M}_{(8)}^{\hat{t}}}^*} = - \frac{8}{9} \hat{s}^2 & & \overline{\mathcal{M}_{(8)}^{\hat{u}} {\mathcal{M}_{(1)}^{\hat{t}}}^*} = \overline{\mathcal{M}_{(1)}^{\hat{t}} {\mathcal{M}_{(8)}^{\hat{u}}}^*} = - \frac{8}{9} \hat{s}^2 . 
\end{align}
To obtain the final result for the different cross sections one needs to combine the pre-factors from equation \eqref{eq:dijetxsec:amplitudes} with the results from equation \eqref{eq:dijetxsec:sqmatrix} and insert them into equation \eqref{eq:dijetxsec:xsec}.

\paragraph{Results} \mbox{} \\ 
For the $q_i q_i \to q_i q_i$ process we then find the following results (with the colour coding \eqred{$t$-channel}, \eqgreen{$u$-channel}, \eqblue{$t$-$u$ channel interference})
\begin{empheq}[box=\eqbox]{align} \label{eq:dijetxsec:results}
	\frac{d \sigma}{d \hat{t}} \Big|_\textrm{QCD} & = \frac{4 \pi \alpha_s^2}{9 \hat{s}^2} \left[ \eqred{\frac{\hat{s}^2 + \hat{u}^2}{\hat{t}^2}} + \eqgreen{\frac{\hat{s}^2 + \hat{t}^2}{\hat{u}^2}} - \eqblue{\frac{2}{3} \frac{\hat{s}^2}{\hat{t} \hat{u}}} \right] \nonumber \\
	\frac{d \sigma}{d \hat{t}} \Big|_{Z'_\textrm{full}}^\textrm{pure} & = \frac{2 \pi \alpha_{Z'}^2}{\hat{s}^2} \left[ \eqred{\frac{\hat{s}^2 + \hat{u}^2}{(\hat{t} - m_{Z'}^2)^2 + \hat{t} \, \Gamma_{Z'}^2 \! \left( \hat{t} \right)}} + \eqgreen{\frac{\hat{s}^2 + \hat{t}^2}{(\hat{u} - m_{Z'}^2)^2 + \hat{u} \, \Gamma_{Z'}^2 \! \left( \hat{u} \right)}} + \eqblue{\frac{2}{3} \hat{s}^2 P \! \left( \hat{t}, \hat{u}, Z' \right)} \right] \nonumber \\
	\frac{d \sigma}{d \hat{t}} \Big|_{Z'_\textrm{full}}^\textrm{int} & = \frac{16 \pi \alpha_s \alpha_{Z'}}{9 \hat{s}^2} \left[ \eqblue{\frac{\hat{s}^2}{\hat{t}} Q \! \left( \hat{u}, Z' \right)} + \eqblue{\frac{\hat{s}^2}{\hat{u}} Q \! \left( \hat{t}, Z' \right)} \right] \nonumber \\
	\frac{d \sigma}{d \hat{t}} \Big|_{Z'_\textrm{eff}}^\textrm{pure} & = \frac{\left| c_{Z'} \right|^2}{2 \pi} \left[ \eqred{\frac{\hat{s}^2 + \hat{u}^2}{\hat{s}^2}} + \eqgreen{\frac{\hat{s}^2 + \hat{t}^2}{\hat{s}^2}} + \eqblue{\frac{2}{3}} \right] \nonumber \\
	\frac{d \sigma}{d \hat{t}} \Big|_{Z'_\textrm{eff}}^\textrm{int} & = \frac{8 \alpha_s \Re \left( c_{Z'} \right)}{9 \hat{s}} \left[ \eqblue{\frac{\hat{s}}{\hat{t}}} + \eqblue{\frac{\hat{s}}{\hat{u}}} \right] \nonumber \\
	\frac{d \sigma}{d \hat{t}} \Big|_{G'_\textrm{full}}^\textrm{pure} & = \frac{4 \pi \alpha_{G'}^2}{9 \hat{s}^2} \left[ \eqred{\frac{\hat{s}^2 + \hat{u}^2}{(\hat{t} - m_{G'}^2)^2 + \hat{t} \, \Gamma_{G'}^2 \! \left( \hat{t} \right)}} + \eqgreen{\frac{\hat{s}^2 + \hat{t}^2}{(\hat{u} - m_{G'}^2)^2 + \hat{u} \, \Gamma_{G'}^2 \! \left( \hat{u} \right)}} - \eqblue{\frac{2}{3} \hat{s}^2 P \! \left( \hat{t}, \hat{u}, G' \right)} \right] \nonumber \\
	\frac{d \sigma}{d \hat{t}} \Big|_{G'_\textrm{full}}^\textrm{int} & = \frac{8 \pi \alpha_s \alpha_{G'}}{9 \hat{s}^2} \left[ \eqred{\frac{\hat{s}^2 + \hat{u}^2}{\hat{t}} Q \! \left( \hat{t}, G' \right)} + \eqgreen{\frac{\hat{s}^2 + \hat{t}^2}{\hat{u}} Q \! \left( \hat{u}, G' \right)} - \eqblue{\frac{1}{3} \left( \frac{\hat{s}^2}{\hat{t}} Q \! \left( \hat{u}, G' \right) + \left\{ \hat{t} \leftrightarrow \hat{u} \right\} \right)} \right] \nonumber \\
	\frac{d \sigma}{d \hat{t}} \Big|_{G'_\textrm{eff}}^\textrm{pure} & = \frac{\left| c_{G'} \right|^2}{9 \pi} \left[ \eqred{\frac{\hat{s}^2 + \hat{u}^2}{\hat{s}^2}} + \eqgreen{\frac{\hat{s}^2 + \hat{t}^2}{\hat{s}^2}} - \eqblue{\frac{2}{3}} \right] \nonumber \\
	\frac{d \sigma}{d \hat{t}} \Big|_{G'_\textrm{eff}}^\textrm{int} & = \frac{4 \alpha_s \Re \left( c_{G'} \right) }{9 \hat{s}} \left[ \eqred{\frac{\hat{s}^2 + \hat{u}^2}{\hat{s} \hat{t}}} + \eqgreen{\frac{\hat{s}^2 + \hat{t}^2}{\hat{s} \hat{u}}} - \eqblue{\frac{1}{3} \frac{\hat{s}}{\hat{t}}} - \eqblue{\frac{1}{3} \frac{\hat{s}}{\hat{u}}} \right] .  
\end{empheq}
In the above equations the functions $P \left( \hat{x}, \hat{y}, X \right)$ and $Q \left( \hat{x}, X \right)$ are defined as
\begin{align}  \label{eq:dijetxsec:widthpropagators}
	P \! \left( \hat{x}, \hat{y}, X \right) & \equiv \frac{\left( \hat{x} - m_X^2 \right) \left( \hat{y} - m_X^2 \right) + \sqrt{\hat{x}} \, \Gamma_X \! \left( \hat{x} \right) \sqrt{\hat{y}} \, \Gamma_X \! \left( \hat{y} \right)}{\left[ \left( \hat{x} - m_X^2 \right)^2 + \hat{x} \, \Gamma_X^2 \! \left( \hat{x} \right) \right] \left[ \left( \hat{y} - m_X^2 \right)^2 + \hat{y} \, \Gamma_X^2 \! \left( \hat{y} \right) \right]} \nonumber \\
	Q \! \left( \hat{x}, X \right) & \equiv \frac{\hat{x} - m_X^2}{\left( \hat{x} - m_X^2 \right)^2 + \hat{x} \, \Gamma_X^2 \! \left( \hat{x} \right)} .
\end{align}
The results for the $q_i q_j \to q_i q_j$ process can be directly read of from equation \eqref{eq:dijetxsec:results} and are given only by the \eqred{$t$-channel} contributions. In equation \eqref{eq:dijetxsec:widthpropagators} the assumption has been made that the combination $\sqrt{\hat{x}} \, \Gamma_X \! \left( \hat{x} \right)$ is real for all values of $\hat{x}$. Equation \eqref{eq:toymodel:dependentwidthnomass} shows that this holds for vanishing quark masses. This is assumed in the numerical calculations as well, since their effect on the differential cross sections is negligible. 

\paragraph{Numerical Calculations} \mbox{} \\
The analytical results derived in this section have to be transformed from partonic dijet cross sections to realistic angular distributions at the LHC. This has been done using the Mathematica package of the MSTW 2008 parton density functions \cite{Martin:2009iq}. Furthermore, the integration over angular variables and the extraction of exclusion limits on parameters has been done using Mathematica. A notebook containing all partonic cross sections, the interface with the parton densities and the extraction of limits is available upon request with the author.

\section{Recast Example}
\label{sec:recastexample}
Here we outline the recasting of existing limits from effective operators for the full theory using the original effective operator bound and the quantified deviation between the full and effective theory. As an example the heavy gluon resonance $\rho$ in a model with right-handed partial compositeness is used \cite{Redi:2013eaa}. When the $\rho$ is integrated out, the effective operator
\begin{empheq}[box=\eqbox]{equation} \label{eq:recastexample:rhcompoerator}
	- \frac{g_\rho^2}{6 m_\rho^2} \sin^4 \phi \left( \bar{q} \gamma^\mu q \right) \left( \bar{q} \gamma^\mu q \right) 
\end{empheq}
is obtained and was used to constrain the parameter space in the $m_\rho$ versus $\sin \phi$ plane\footnote{Here we use the simplification of removing the handedness of the Standard Model quarks.}. With the use of this example we outline the steps needed to rescale this limit to include the full theory effects. 

\begin{figure}[!ht]
	\centering
	\includegraphics[width=0.6\textwidth]{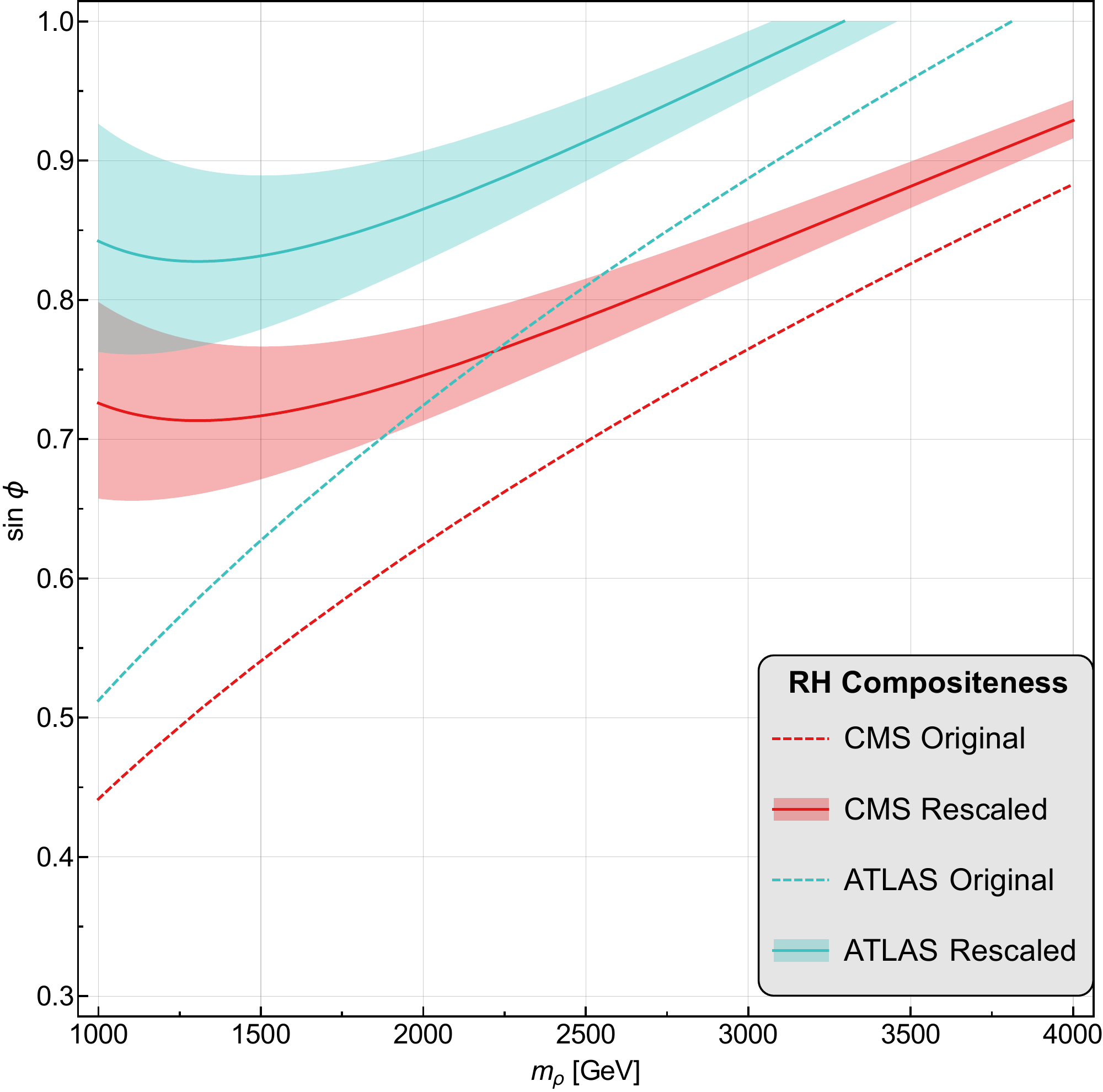} 
	\caption{Recasting of existing effective operator limits using the quantified deviation between effective and full theory for the example of partial right-handed compositeness \cite{Redi:2013eaa}. A detailed description of the procedure is provided in the text. The dashed lines represent the naive limits from effective operator constraints, whereas the solid lines show the more reliable rescaled limits. The theoretical errors introduced by the rescaling procedure are shown by the bands around the solid lines.}
	\label{fig:recastexample:rhcomprecast}
\end{figure}

\begin{enumerate}
	\item 
		The first step is to obtain the experimental limit on either one of the toy models using the effective operator coefficient and compare with the experimental limit
		\begin{equation} \label{eq:recastexample:explimit}
			\left| c \right| = \left| - \frac{g^2}{2 m^2} \right| = \frac{2 \pi}{\Lambda_\textrm{exp}^2} .
		\end{equation}
		In this case this is the $Z'$ operator and it establishes the exclusion contour in the mass versus coupling plane.
	\item 
		Convert the exclusion contour to a limit on the coupling $g$ as a function of the mass $m$ and then use equation \eqref{eq:results:deviationfit} with the fitted parameter $C_{Z'}$ from equation \eqref{eq:results:cfitfchi} to rescale the exclusion limit. A realistic limit on the full theory behind the $Z'$ toy model using the experimental limit is then obtained.
	\item 
		Compare the effective operator coefficients and express the parameters of the model under consideration in terms of the toy model parameters. For the example at hand we obtain
		\begin{equation} \label{eq:recastexample:couplingid}
			g = \sqrt{\tfrac{1}{3}} g_\rho \sin^2 \phi ,
		\end{equation}
		where $m$ equals $m_\rho$ by definition and drops out. 
	\item
		Express the exclusion limits on the toy model in terms of the model parameters using equation \eqref{eq:recastexample:couplingid} to obtain realistic exclusion limits for the considered model. For the model considered the limits are expressed in the mass versus $\sin \phi$ plane, using $g_\rho = 3$ for the identification.
\end{enumerate}
Following these steps for the model with right-handed compositeness results for the adjusted exclusion limits are presented in figure \ref{fig:recastexample:rhcomprecast}. We observe that the exclusion limits are significantly reduced\footnote{The deviation between full and effective theory limits has been obtained based on the ATLAS analysis and has also been applied to the CMS limits. Therefore, the rescaled limits should be seen as an indication and a more detailed analysis of the deviation is required.}. However, it is noted that the exclusion limits quoted in reference \cite{Redi:2013eaa} remain unchanged due to overlap between the excluded regions from effective operators and dijet resonance searches.

\section{Results for LHC14}
\label{sec:resultslhc14}

\begin{figure}[!ht]
	\centering
	\includegraphics[width=0.46\textwidth]{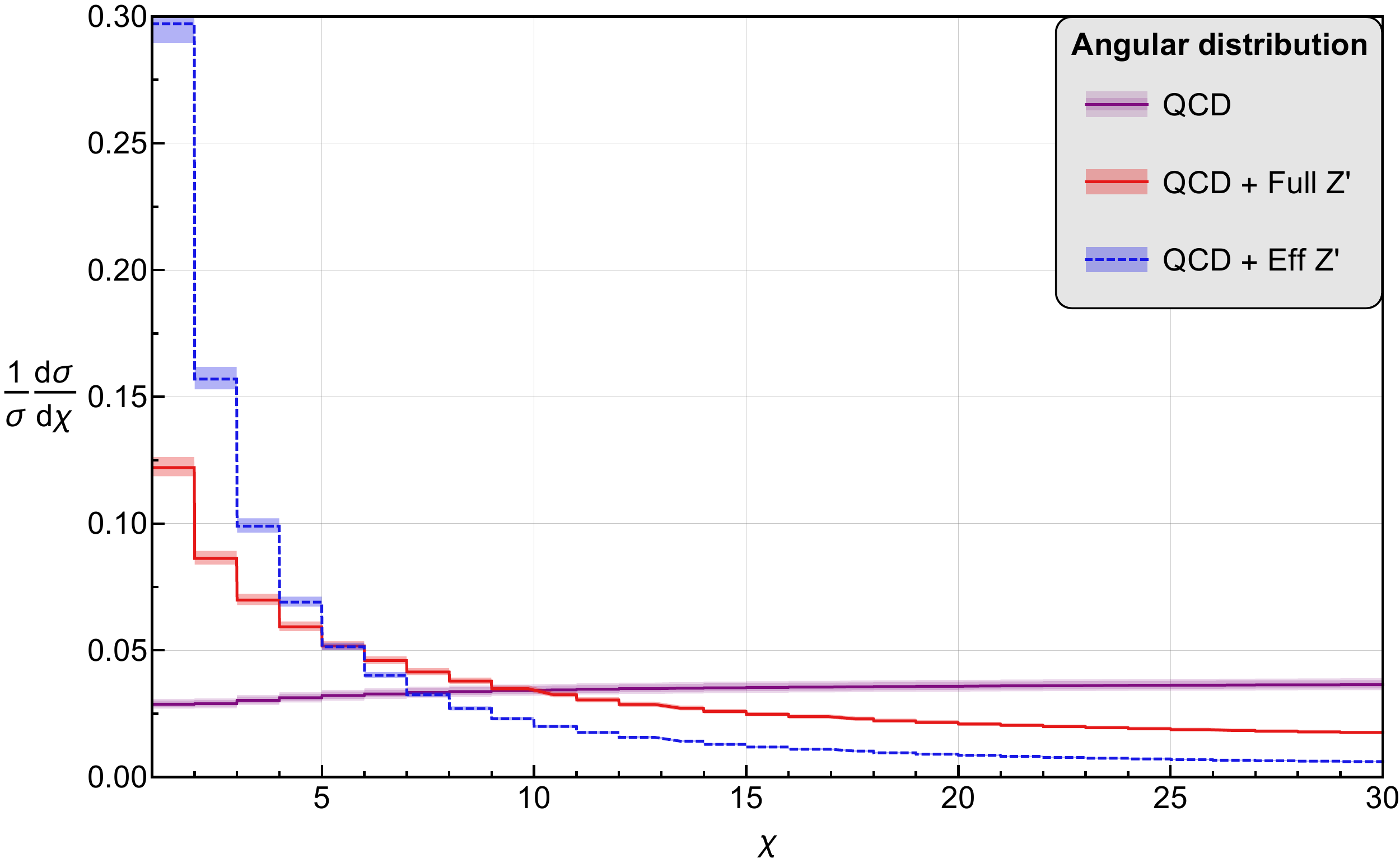} \hspace{2mm}  
	\includegraphics[width=0.46\textwidth]{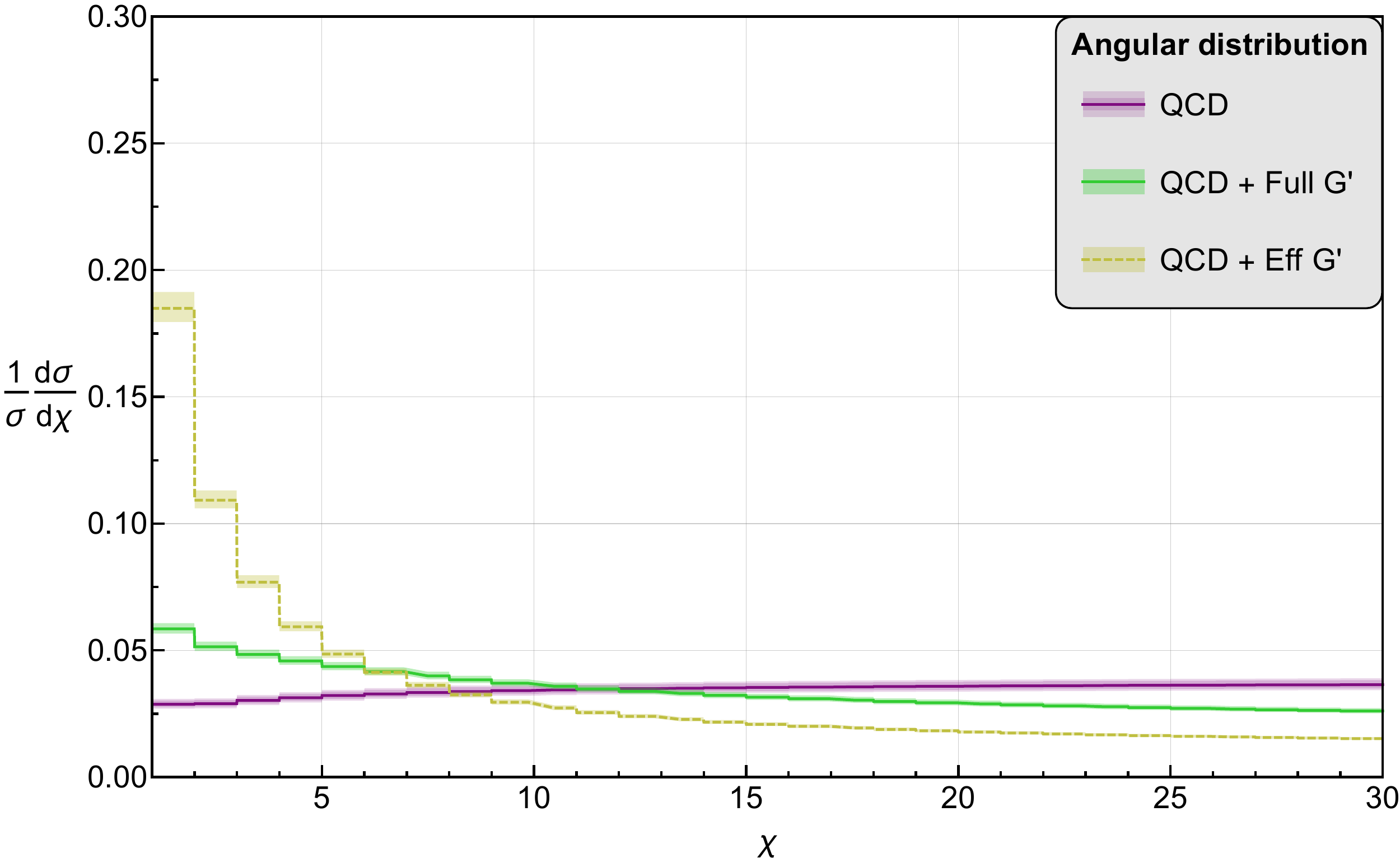} 
	\caption{Equivalent of figure \ref{fig:analysis:dsigmdchi} for the $14 \; \textrm{TeV}$ LHC run with an integrated luminosity of $100 \; \textrm{fb}^{-1}$. This distribution has been obtained for a centre of mass energy integration from $\sqrt{\hat{s}_\textrm{min}} = 4 \; \textrm{TeV}$ to $\sqrt{\hat{s}_\textrm{max}} = 7 \; \textrm{TeV}$.}
	\label{fig:resultslhc14:dsigmdchi}
\end{figure}

\begin{figure}[!ht]
	\centering
	\includegraphics[width=0.46\textwidth]{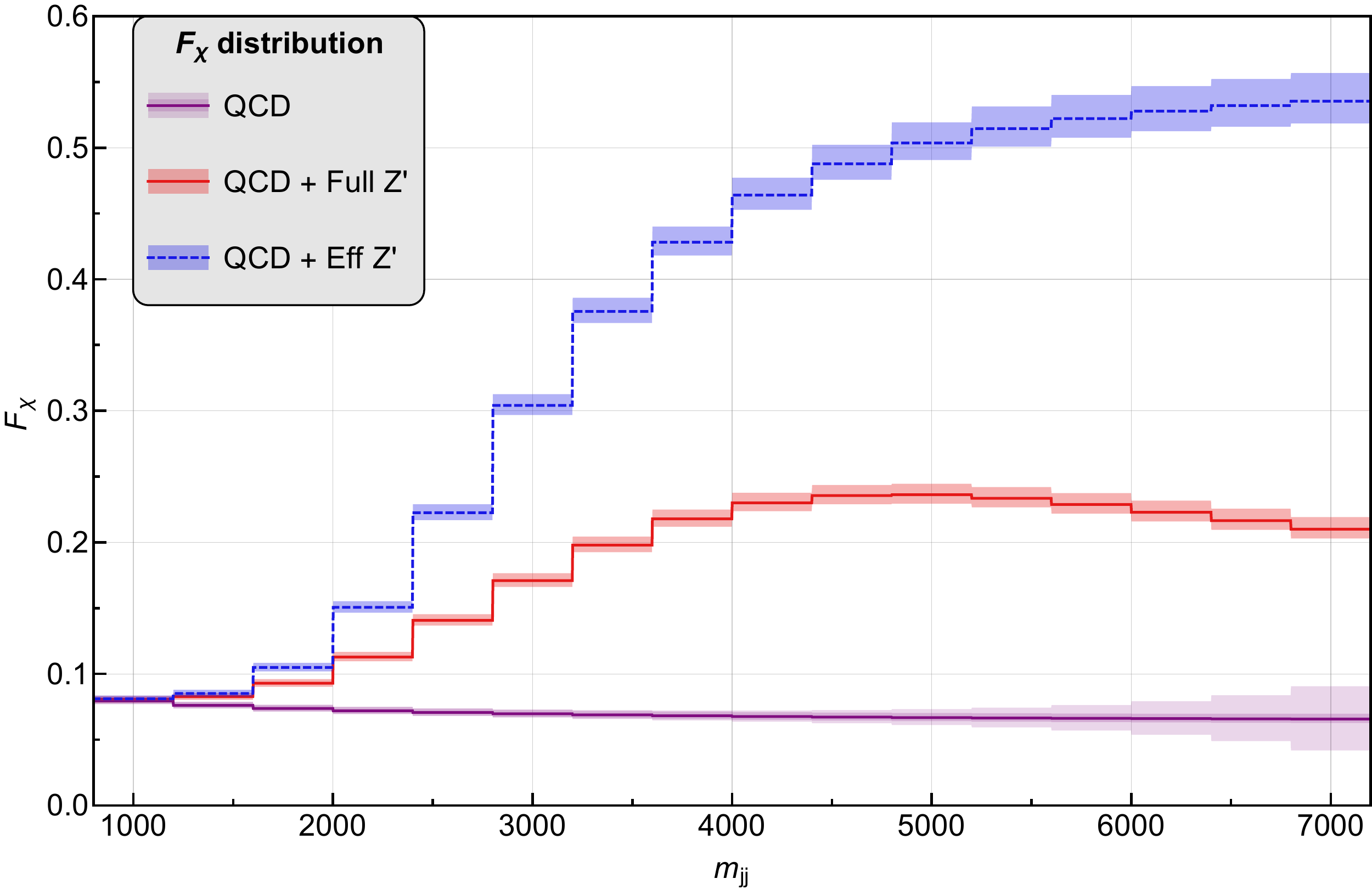}  \hspace{2mm}  
	\includegraphics[width=0.46\textwidth]{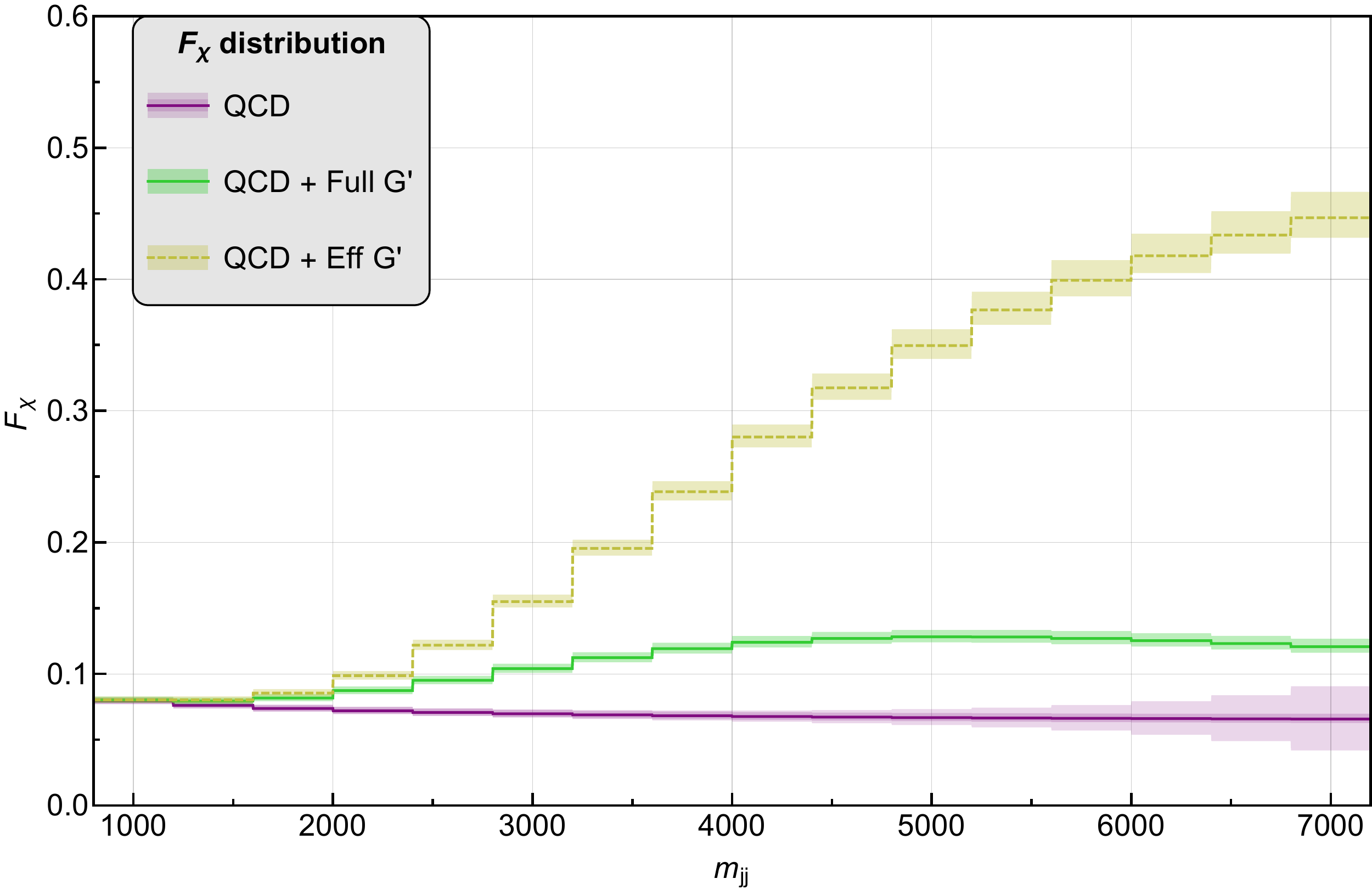} 
	\caption{Equivalent of figure \ref{fig:analysis:fchi} for the $14 \; \textrm{TeV}$ LHC run with an integrated luminosity of $100 \; \textrm{fb}^{-1}$.}
	\label{fig:resultslhc14:fchi}
\end{figure}

The results presented in the main body of this work all have been obtained for the LHC operating at a centre of mass energy of $7 \; \mathrm{TeV}$ with an integrated luminosity of $5 \; \textrm{fb}^{-1}$. However, it is even more interesting to see the effects at a centre of mass energy of $14 \; \textrm{TeV}$, since the partonic centre of mass energy significantly increases. With a higher partonic centre of mass energy the average transfer energy will increase and the effective expansion will be less reliable for the same points in the full theory parameter space. On the other hand, when the Large Hadron Collider operates at $14 \; \textrm{TeV}$ it will gather more data, resulting in an increased integrated luminosity and producing more precise results. For this purpose the results presented in this appendix are based on an integrated luminosity of $100 \; \textrm{fb}^{-1}$ for the $14 \; \textrm{TeV}$ run. In figures \ref{fig:resultslhc14:dsigmdchi} and \ref{fig:resultslhc14:fchi} the angular distributions used in the CMS and ATLAS experiments are presented.

\begin{figure}[!ht]
	\centering
	\includegraphics[width=0.6\textwidth]{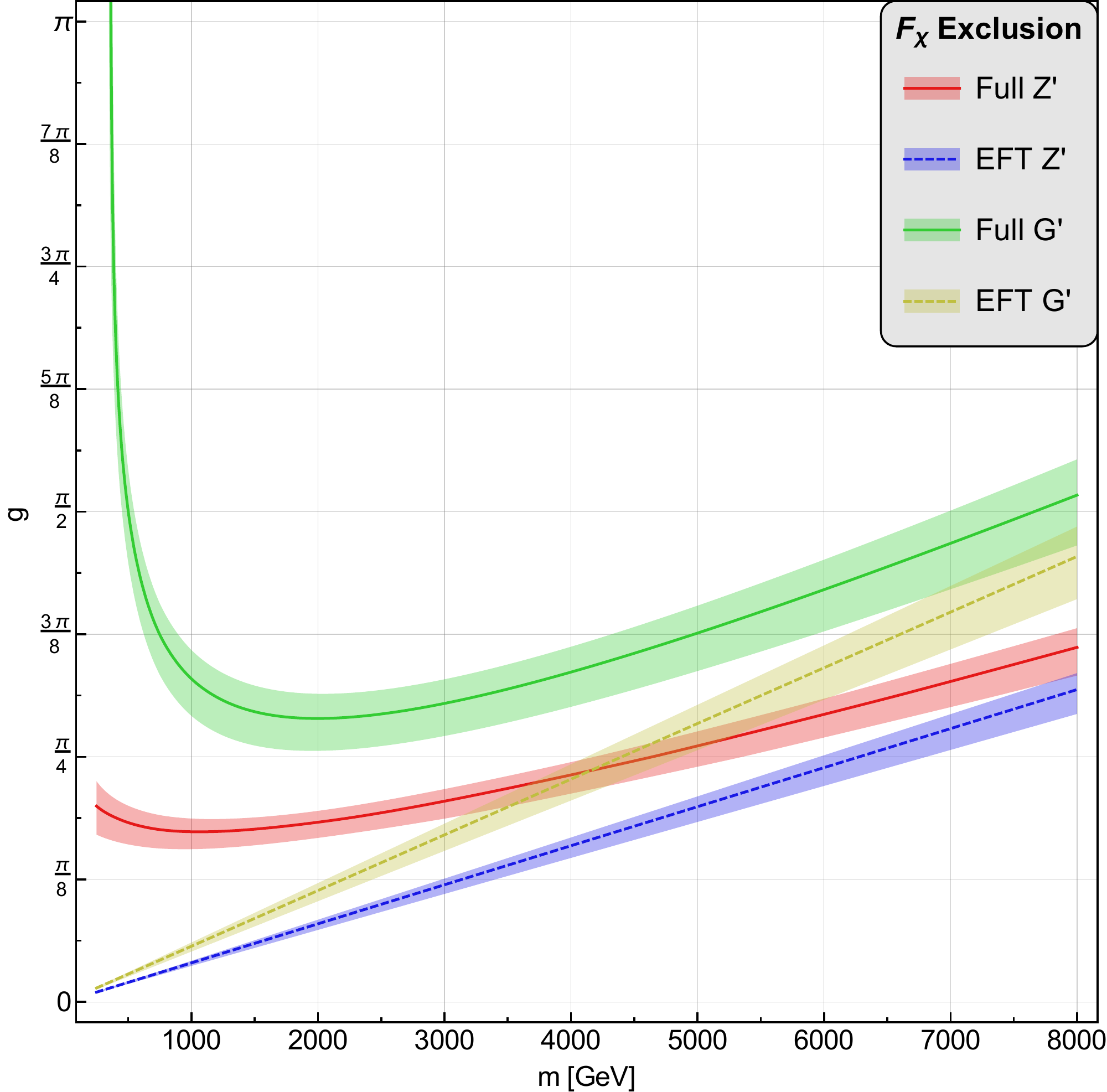} 
	\caption{Equivalent of figure \ref{fig:results:fchiexclusion} for the $14 \; \textrm{TeV}$ LHC run with an integrated luminosity of $100 \; \textrm{fb}^{-1}$.}
	\label{fig:resultslhc14:fchiexclusion}
\end{figure}

The theoretical measure based on the $F_\chi$ distribution from section \ref{sec:results} needs to be modified. Based on figure \ref{fig:resultslhc14:fchi} the binning for the $\chi^2$ analysis is changed to $\sqrt{\hat{s}}$ ranging from $2000 \; \textrm{GeV}$ to $7200 \; \textrm{GeV}$ with steps of $400 \; \textrm{GeV}$. This allows for a more thorough scanning of the full kinematic reach of the $14 \; \textrm{TeV}$ LHC run. The resulting exclusion limits for the full and effective descriptions of the toy models are presented in figure \ref{fig:resultslhc14:fchiexclusion}. The limits in figure \ref{fig:resultslhc14:fchiexclusion} correspond to values
\begin{equation} \label{eq:resultslhc14:oplimit}
	\Lambda_{Z'} = 28.3^{+2.4}_{-1.4} \; \textrm{TeV} , \qquad \qquad \Lambda_{G'} = 19.9^{+2.1}_{-1.2} \; \textrm{TeV} .
\end{equation}

\begin{figure}[!ht]
	\centering
	\includegraphics[width=0.6\textwidth]{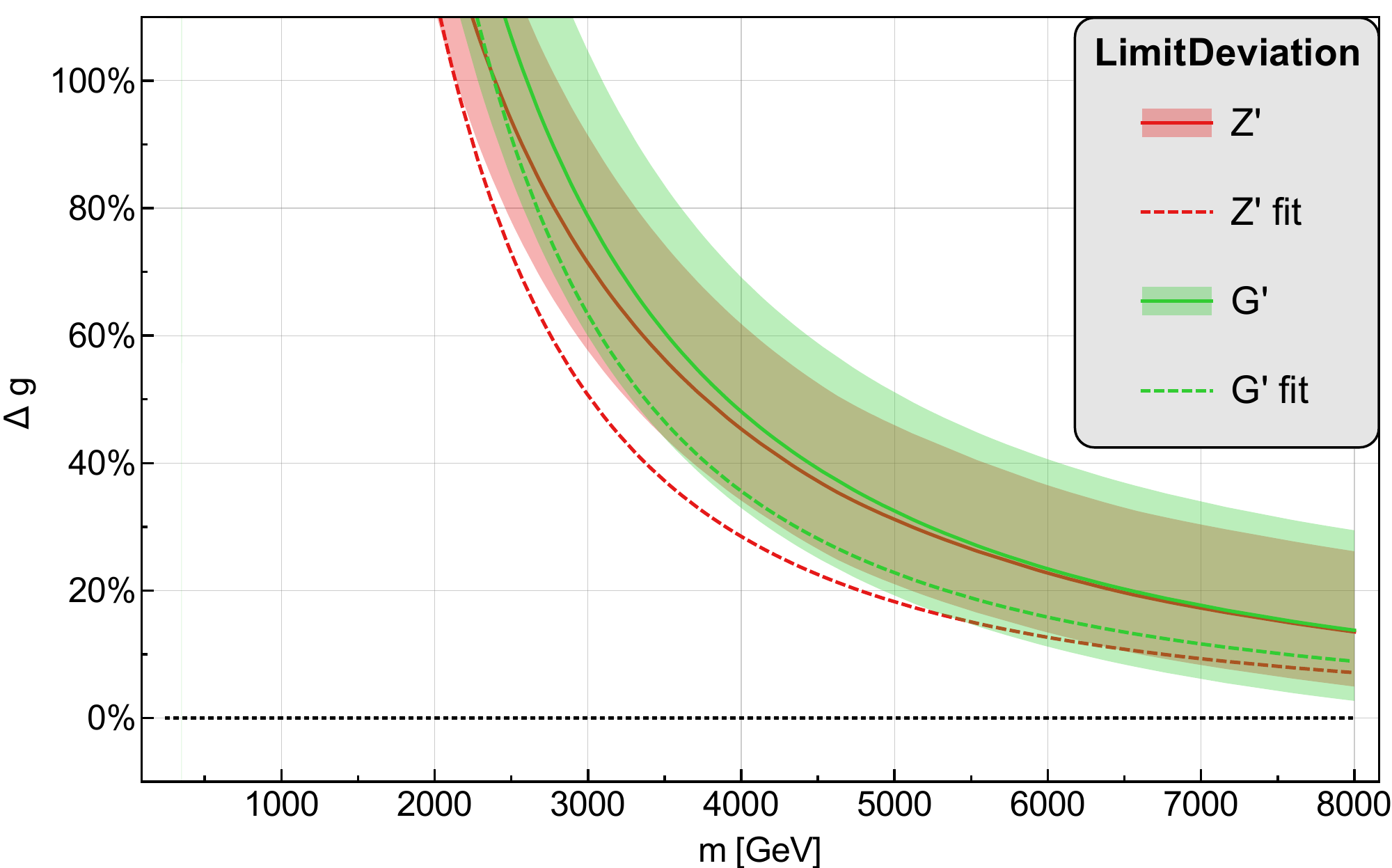} 
	\caption{Equivalent of figure \ref{fig:results:deviation} for the $14 \; \textrm{TeV}$ LHC run with an integrated luminosity of $100 \; \textrm{fb}^{-1}$.}
	\label{fig:resultslhc14:deviation}
\end{figure}

The deviation between the full and the effective descriptions is given in figure \ref{fig:resultslhc14:deviation}. The fitted values for the $C$ parameter in this case are given by
\begin{equation} \label{eq:resultslhc14:cfitfchi}
	C_{Z'} = 2.14^{+0.21}_{-0.20} \; \textrm{TeV} , \qquad C_{G'} = 2.39^{+0.27}_{-0.22} \; \textrm{TeV} .
\end{equation}
As expected, we observe that the deviation is larger for any chosen mass of the particle in the full theory compared to the $7 \; \textrm{TeV}$ result from figure \ref{fig:results:deviation}. This is explained by the higher average transfer energies, which are presented in table \ref{tab:resultslhc14:transferenergies}. Therefore, one should be even more careful when extracting limits on BSM models from four-quark effective operator bounds when using $14 \; \textrm{TeV}$ data.

\begin{table}[!ht]
	\centering
	\begin{tabular}{| c | c | c | c c | c c |}
		\toprule[1pt]
		region & $\sqrt{\left| \left\langle q^2 \right\rangle \right|}$ & QCD & Full $Z'$ & Eff $Z'$ & Full $G'$ & Eff $G'$ \\
		\midrule[1pt]
		\multirow{3}{*}{total} & $\sqrt{\left| \left\langle \hat{s} \right\rangle \right|}$ & $2.42^{+0.24}_{-0.21}$ & $2.52^{+0.23}_{-0.20}$ & $2.78^{+0.23}_{-0.20}$ & $2.48^{+0.24}_{-0.20}$ & $2.55^{+0.24}_{-0.20}$ \\
		& $\sqrt{\left| \left\langle \hat{\scalebox{0.88}{$t$}} \right\rangle \right|}$ & $0.73^{+0.07}_{-0.06}$ & $0.87^{+0.08}_{-0.06}$ & $1.15^{+0.09}_{-0.08}$ & $0.79^{+0.07}_{-0.06}$ & $0.88^{+0.08}_{-0.07}$ \\
		& $\sqrt{\left| \left\langle \hat{u} \right\rangle \right|}$ & $2.31^{+0.23}_{-0.20}$ & $2.36^{+0.22}_{-0.18}$ & $2.53^{+0.21}_{-0.18}$ & $2.35^{+0.22}_{-0.19}$ & $2.39^{+0.22}_{-0.19}$ \\
		\midrule[1pt]
		\multirow{3}{*}{central} & $\sqrt{\left| \left\langle \hat{s} \right\rangle \right|}$ & $2.42^{+0.25}_{-0.21}$ & $2.66^{+0.20}_{-0.17}$ & $3.17^{+0.21}_{-0.18}$ & $2.53^{+0.22}_{-0.19}$ & $2.81^{+0.23}_{-0.19}$ \\
		& $\sqrt{\left| \left\langle \hat{\scalebox{0.88}{$t$}} \right\rangle \right|}$ & $1.39^{+0.14}_{-0.12}$ & $1.56^{+0.12}_{-0.10}$ & $1.90^{+0.13}_{-0.11}$ & $1.47^{+0.13}_{-0.11}$ & $1.66^{+0.13}_{-0.11}$ \\
		& $\sqrt{\left| \left\langle \hat{u} \right\rangle \right|}$ & $1.97^{+0.20}_{-0.17}$ & $2.15^{+0.16}_{-0.14}$ & $2.54^{+0.17}_{-0.15}$ & $2.06^{+0.18}_{-0.15}$ & $2.27^{+0.18}_{-0.15}$ \\
		\bottomrule[1pt]
	\end{tabular}
	\caption{Equivalent of table \ref{tab:results:transferenergies} for the $14 \; \textrm{TeV}$ LHC run with an integrated luminosity of $100 \; \textrm{fb}^{-1}$.}
	\label{tab:resultslhc14:transferenergies}
\end{table}

\bibliographystyle{JHEP}
\bibliography{hceff}

\end{document}